\documentclass[10pt,twocolumn,superscriptaddress,nofootinbib]{revtex4-1}
\usepackage[total={7.0in,9.5in}, top=1.0in, includefoot]{geometry}
\usepackage{verbatim}
\usepackage{epsfig}
\usepackage{subfigure}
\usepackage{amsmath}
\usepackage[usenames,dvipsnames]{color}
\usepackage{amssymb}
\usepackage{setspace}
\usepackage{graphicx} 
\usepackage{times}
\usepackage{amsthm}
\usepackage{hyperref}
\usepackage{titlesec}
\usepackage{array}
\usepackage{siunitx}
\usepackage{caption}

\hypersetup{bookmarks=true, unicode=false, pdftoolbar=true, pdfmenubar=true, pdffitwindow=false, pdfstartview={FitH}, pdfcreator={Daniel Larremore}, pdfproducer={Daniel Larremore}, pdfkeywords={} {} {}, pdfnewwindow=true, colorlinks=true, linkcolor=red, citecolor=Green, filecolor=magenta, urlcolor=cyan,}

\renewcommand\labelenumi{(\roman{enumi})}
\renewcommand\theenumi\labelenumi

\renewcommand\figurename{Fig.}

\DeclareCaptionFont{xipt}{\fontsize{10}{11}\mdseries}
\usepackage[font=xipt,labelfont=bf]{caption}
\titlespacing\section{0pt}{4pt plus 1pt minus 1pt}{1pt plus 1pt minus 1pt}

\titlespacing\subsection{0pt}{4pt plus 1pt minus 1pt}{2pt plus 1pt minus 1pt}

\titleformat{\section}[block]{\bfseries\raggedright\large}{\thesection\hspace{2pt}.}{4pt}{}

\titleformat{\subsection}[block]{\bfseries\raggedright\mdseries}{\thesection\hspace{2pt}\thesubsection.}{4pt}{}

\begin{document}

\author{Aniko Hannak}
	\email{hannak@ifi.uzh.ch}
	\affiliation{\vspace{-3pt}Department of Computer Science, University of Z{\"u}rich, Z{\"u}rich, Switzerland}
	\thanks{The four authors contributed equally to this work. Author order is random.}
\author{Kenneth Joseph}
	\email{kjoseph@buffalo.edu}
	\affiliation{\vspace{-3pt}Department of Computer Science and Engineering, University at Buffalo, SUNY, Buffalo, NY, USA}	
	\thanks{The four authors contributed equally to this work. Author order is random.}
\author{Andrei Cimpian}
	\email{andrei.cimpian@nyu.edu }
	\affiliation{\vspace{-3pt}Department of Psychology, New York University, New York, NY, USA}
	\thanks{The four authors contributed equally to this work. Author order is random.}
\author{Daniel B. Larremore}
	\email{daniel.larremore@colorado.edu}
	\affiliation{\vspace{-3pt} Department of Computer Science, University of Colorado, Boulder, CO, USA}
	\affiliation{\vspace{-3pt}BioFrontiers Institute, University of Colorado, Boulder, CO, USA}
	\thanks{The four authors contributed equally to this work. Author order is random.}
	
\title{Explaining Gender Differences in Academics' Career Trajectories}
\begin{abstract}
\vspace{0.25cm}
\noindent Academic fields exhibit substantial levels of gender segregation. To date, most attempts to explain this persistent global phenomenon have relied on limited cross-sections of data from specific countries, fields, or career stages. Here we used a global longitudinal dataset assembled from profiles on ORCID.org to investigate which characteristics of a field predict gender differences among the academics who leave and join that field. Only two field characteristics consistently predicted such differences: (1) the extent to which a field values raw intellectual talent (``brilliance'') and (2) whether a field is in Science, Technology, Engineering, and Mathematics (STEM). Women more than men moved \textit{away from} brilliance-oriented and STEM fields, and men more than women moved \textit{toward} these fields. Our findings suggest that stereotypes associating brilliance and other STEM-relevant traits with men more than women play a key role in maintaining gender segregation across academia.
\vspace{0.25cm}
\end{abstract}
\maketitle

Gender segregation in academia---and the workplace more generally---remains substantial well into the 21st century~\cite{levanon2016persistence,cech2013self,england2010gender,NSF2017earned, NSF2017recipients, UNESCO2019}, undermining gender equity in earnings and status~\cite{hegewisch2010separate,blau2007gender,blau2017gender}. Although understanding the causes of this phenomenon, especially as it concerns STEM fields, is a priority for many governmental and international agencies~\cite{UNwomen,NSF2019broadening,fawcett2019}, a clear view is complicated by several factors. First, the reasons why women and men differentially leave or join certain fields are best understood in the context of their broader career trajectories---what they did after leaving the field in question or before joining it~\cite{jacobs1989revolving}. Yet, most analyses of gender segregation in academia to date have not examined individual-level longitudinal information of this sort. Second, academic fields vary considerably in their levels of gender segregation, regardless of whether they are in STEM, the social sciences, or the humanities~\cite{cheryan2017some,leslie2015expectations}. Although the issue of women’s underrepresentation in STEM has received substantial attention, few investigations to date have considered a wide enough range of fields to be able to identify cross-cutting explanatory factors underlying gender segregation across academia---or heterogeneity in such factors (for some important exceptions, see~\cite{Ganley2018bias,ceci2014women,miller2015bachelor,leslie2015expectations}). Third, there is systematic variability in the gender composition of academic fields across countries~\cite{UNESCO2019,breda2018societal,charles2009indulging}, yet investigations of this phenomenon have predominantly focused on a specific cultural context, potentially missing explanatory factors that emerge at broader levels of analysis (for some important exceptions, see \cite{charles2005occupational,charles2009indulging, miller2015women,breda2018societal}). Fourth, there are secular trends in the extent to which men and women differ in the skills needed for various careers~\cite{hyde2009gender} and are subject to stereotypes relevant to these careers~\cite{charlesworth2019patterns,eagly2019gender}, as well as in other factors that may contribute to gender segregation~\cite{breda2018societal}. Thus, analyses of this phenomenon are most informative at a broad temporal scale, which creates additional challenges---most notably, data availability. 

Here we provide the first investigation of gender segregation in academia that simultaneously satisfies all of the above criteria, in that (i) it takes into account individuals’ movements between fields; (ii) it encompasses as many as 30 fields across STEM, social sciences, and the humanities; and it includes information on individuals (iii) from over 200 different countries (iv) across more than 6 decades.

To accomplish this goal, we created a unique dataset, which we have now made freely available. This dataset was compiled from two independent sources. One source consisted of publicly available author profiles from ORCID.org (Open Researcher and Contributor ID), a not-for-profit organization that maintains a global database of scholars, their educational and employment history, and their published research (see Section~\ref{section:ORCID} in the Supplementary Text and Supplementary Fig.~\ref{Fig:ORCIDannotations}). The ORCID profiles allowed us to comprehensively compare how women and men move between fields.\footnote{Throughout, we use the terms {\it move}, {\it switch}, and {\it transition} interchangeably to refer to an observed change in an ORCID user's academic field.} The second source of data consisted of a survey of academics from 30 different fields, in which they rated their own fields along several dimensions \cite{leslie2015expectations}. When combined with the ORCID profiles, these data offered unprecedented insight into the processes underlying gender segregation in academia.

We seek to understand gender segregation in academia by explaining how and why academics move between fields. This approach provides an update to the common metaphor of a ``leaky pipeline.'' Most pipeline analyses compare the proportion of women (or men) at consecutive stages in the professional trajectory of a field’s members (e.g., bachelor’s degrees vs. PhD degrees;~\cite{ceci2014women, miller2015bachelor}). In these analyses, a field’s pipeline is said to be leaking women (or men) if the proportion of women (or men) in the field declines from one career stage to the next. Although useful, this approach is intrinsically limited by the fact that it cannot provide insight into why gaps emerge when they do. For instance, are more women than men leaving the field, or more men than women joining it, or both? Where did the women who left go, and where did the men who joined come from? What is it about a field that explains gender-differentiated career transitions into and out of it? Traditional pipeline analyses cannot answer questions such as these, which are essential for an adequate understanding of gender gaps in representation. Our approach, which engages with the complexities of the ``branching pipeline'' of women’s and men’s career trajectories~\cite{fuhrmann2011improving}, may provide a promising step toward this deeper theoretical understanding.

With the extensive new dataset we created by enriching the ORCID data with field attributes (as rated by academics in these fields;~\cite{leslie2015expectations}), we were able to compare five distinct explanations for why women and men in academia might follow different paths. We focused on these explanations, which we describe in the next section, for the same reasons they were included in Leslie, Cimpian, Meyer, and Freeland's survey of academics~\cite{leslie2015expectations}: Although by no means exhaustive, they represent some of the more prominent and well-supported theories in the literature. They also represent a range of perspectives on the mechanisms underlying gender segregation, from those that emphasize the culture of a field and gender stereotypes to those that focus on hypothesized differences between women's and men's preferences and abilities. Comparing multiple theoretical perspectives with the same data and methods, rather than focusing narrowly on just one (type of) perspective, is arguably more likely to lead to theoretical progress.

\noindent{\bf{Five potential explanations.}}
One explanation for gender segregation in academia appeals to differences among fields in the extent to which their members believe that success depends on innate intellectual ability (``brilliance''). Because cultural stereotypes associate men more than women with this trait \cite{bian2017gender,bian2018evidence,storage2020adults}, fields that value brilliance---which include many STEM fields---may be more welcoming to men than women. In fact, several studies (focusing mostly on U.S. bachelor’s and PhD degrees) have shown that the brilliance orientation of a field is negatively associated with the proportion of women among degree recipients, even when holding constant a number of other relevant factors \cite{leslie2015expectations, storage2016frequency, ito2018factors, meyer2015women}. This hypothesis predicts that women should be relatively more likely than men to transition toward academic fields with lower brilliance orientations and, conversely, that men should be more likely than women to transition toward academic fields with higher brilliance orientations (see Supplementary Table~\ref{Tab:survey} for the measure of a field's brilliance orientation).

A second explanation appeals to differences among academic fields in the extent to which succeeding in them is compatible with work-life balance. More women than men report that they value flexibility in work schedules and achieving some level of work-life balance~\cite{mccabe2019shines,hakim2006women}, so fields that require longer hours---particularly on-campus hours, which are less flexible---may be less welcoming to women’s participation and more welcoming to men's. This hypothesis predicts that women should be relatively more likely than men to transition toward academic fields with lower on-campus workloads and, conversely, that men should be more likely than women to transition toward academic fields with higher workloads (see Supplementary Table~\ref{Tab:survey} for the measure of a field's on-campus workload).

A third explanation appeals to differences among academic fields in the extent to which they focus on inanimate objects vs. living things, including people. Prior work has suggested that men tend to prefer occupations that focus on inanimate objects more than women do, whereas women tend to prefer occupations that deal with living things and people more than men do~\cite{lippa1998gender,su2009men}. A more recent formulation of this idea appeals to the concepts of systemizing (i.e., analyzing the world as a system of inputs and outputs; see also the notion of systematic self-concepts~\cite{cech2013self,lee1998kids}), which is claimed to be more common in men, and empathizing (i.e., intuitively understanding others’ mental states), which is claimed to be more common in women~\cite{baron2002extreme}. From this perspective, the more a field values systemizing relative to empathizing, the less welcoming and/or appealing it should be to women and the more welcoming and/or appealing it should be to men~\cite{billington2007cognitive}. This hypothesis predicts that women should be relatively more likely than men to transition toward academic fields with a weaker emphasis on systemizing (vs. empathizing) and, conversely, that men should be more likely than women to transition toward academic fields with a stronger emphasis on systemizing (vs. empathizing) (see Supplementary Table~\ref{Tab:survey} for the measure of a field's emphasis on systemizing vs. empathizing). 

Fields in which empathizing is valued may also be more likely to facilitate their members' pursuit of communal goals (e.g., working with others, helping others). If so, this test of the systemizing--empathizing hypothesis may also bear on the proposal that women are underrepresented in fields whose pursuit is typically seen as inconsistent with the pursuit of communal goals (such as many fields in STEM;~\cite{diekman2010congruity}). On the assumption that the systemizing--empathizing variable can serve as a proxy for a field's compatibility with communal goals, we would again expect women (more than men) to transition toward fields with a stronger emphasis on empathizing.

A fourth explanation appeals to differences among academic fields in their selectivity. Even when women and men do not differ on average with respect to a certain trait or ability (e.g., intelligence), some have suggested that differences in variability may still exist, with men being overrepresented at both the high and low ends of the relevant distributions (e.g., ~\cite{hedges1995sex,karwowski2016greater,makel2016sex,baye2016gender}; but see~\cite{feingold1994gender,odea2018gender}). Thus, the more selective a field is, the more likely it is to recruit individuals from the extreme high end of the relevant ability distributions, and as a result the bigger the gender gaps favoring men should be (because, on this argument, men are increasingly overrepresented relative to women as one approaches the tails). This hypothesis predicts that women should be relatively more likely than men to transition toward less selective academic fields and, conversely, that men should be more likely than women to transition toward more selective academic fields (see Supplementary Table~\ref{Tab:survey} for the measure of a field's selectivity). 

This test of the selectivity hypothesis relies on several assumptions. First, it assumes that existing members of a field select new members based exclusively on their abilities and thus choose higher-ability candidates when the field is more selective. Second, it assumes that the pools of candidates available across fields are roughly matched in terms of the means and distributions of the relevant abilities. Without these two assumptions, higher selectivity would not necessarily translate into more right-tail selections. Third, this test of the selectivity hypothesis assumes that men are more variable than women in all or most abilities that are relevant to success in the 30 fields under consideration. While this assumption is debatable (as are the other two), it is nevertheless informative to investigate whether selectivity relates to observed patterns of gender segregation. For instance, if we found that segregation does not track selectivity, then no matter where the science ultimately settles with respect to the claim of greater male variability, we would know that this variability does not have much of a bearing on academics' career transitions. 

Finally, a fifth explanation appeals to differences in the environments of STEM fields vs. fields outside of STEM that are not reducible to the attributes described above but that nevertheless make STEM fields less welcoming or appealing to women than men. This hypothesis is motivated by the long history of women's underrepresentation in (many) STEM fields, which persists into the present and is the focus of concerted research and policy-making efforts internationally~\cite{cheryan2017some,ceci2014women,moss2012science,UNwomen}. This hypothesis predicts that women should be relatively more likely than men to transition toward non-STEM fields and, conversely, that men should be more likely than women to transition toward STEM fields.

\section*{Methods}

\noindent{\bf{Identifying career transitions: The ORCID dataset.}}
We tested the explanations above using a new dataset created from public profiles on ORCID, which were augmented with field characteristics obtained from a survey of academics \cite{leslie2015expectations}. Since ORCID does not collect field, career status, or gender information from its users, we had to infer these metadata. We processed ORCID data in three steps: (i) cleaning the data; (ii) inferring the roles, fields, and likely perceived gender of each ORCID user; and (iii) identifying field transitions (see Section \ref{section:ORCID} in the Supplementary Text).

Starting with an initial 6,485,785 unique ORCID users with 5,307,437 affiliations, we first removed (i) users without a first and last name, which we needed to estimate an association with gender; (ii) affiliations that did not include a department name or equivalent, which we needed to infer a user's academic field; or (iii) affiliations that lacked either a position/role (e.g., ``bachelor's degree,'' ``postdoc'') or an associated date, which we needed to infer career transitions. These filtering steps resulted in 3,988,331 remaining affiliations from 1,287,228 users.

We inferred roles, fields, and cultural name--gender associations using three distinct algorithms. First, a {\it role} was assigned to each affiliation from the following list of roles: bachelor's, master's/postgraduate, PhD, postdoc, professor/department head, or unknown (see Section \ref{sub:stages} in the Supplementary Text). The mapping between these roles’ various aliases and names in other languages was done by recursively accumulating a list of hand-checked aliases used in regular expressions. 

Second, a {\it field} was assigned to each affiliation using a rule-based matching algorithm (see Section \ref{sub:fields} in the Supplementary Text). We discarded any affiliation that had (i) no matching field, (ii) two or more matching fields, or (iii) one matching field that was not among the list of 30 fields surveyed by Leslie, Cimpian, and colleagues \cite{leslie2015expectations}. This conservative approach resulted in 1,274,089 affiliations from 685,649 users. Each affiliation was also labeled with a {\it geographic region}, based on the classifications provided by the United Nations Statistics Division~\cite{UNgeography} (see Section \ref{sub:region} in the Supplementary Text).

Third, we inferred {\it name--gender associations} using a cultural consensus model~\cite{batchelder1988test} that computed the Bayesian posterior probability that a person’s name was culturally understood to belong to a woman (or complementarily, a man) based on data from 44 different sources, ranging from the U.S. Social Security Administration’s names database to a list of the world’s Olympic Athletes (see Section~\ref{sub:genders} in the Supplementary Text). Names that did not appear in any of the 44 reference datasets were submitted to Genni~\cite{torvik2016ethnea}, a service that takes into account the perceived ethnicity of first and last names to improve estimates of gender from first names. Finally, names with posterior probabilities or Genni scores of $\geq 0.9$ or $\leq 0.1$ were labeled as being likely to be associated with a woman or a man, respectively. Names with scores between 0.1 and 0.9 were not included in our analyses (20.5\% of names).  We were able to make name--gender associations for 550,961 of the 685,649 users with at least one affiliation linked to an academic field in our survey data, resulting in 1,027,250 affiliations from 550,961 people. 

As a reliability check, 600 ORCID profiles were chosen uniformly at random and provided to both the cultural consensus model and a panel of research assistants who coded perceived gender, via pronoun usage and photographs, based on a web search of individuals' names and their recent institutional employment. The gender ratios in this sample were indistinguishable between the model and the research assistants (37.87\% and 37.95\% women) with disagreement on only 1.7\% of coded individuals.\footnote{We acknowledge that it is impossible to determine the gender of any {\it{individual}} person using this method. Rather, the application of gendered labels to ORCID identifiers represents an aggregate probability that a given name will be culturally perceived to match a binary gender. Although we use ``men'' and ``women'' as shorthand to describe this aggregate probability in our manuscript, these labels should only be used in aggregate as they may misrepresent the gender of any given individual.}

We identified {\it field transitions} among the 1,027,250 affiliations by sorting each individual’s affiliations by date whenever possible, or by role sequence when no dates were provided (see Section \ref{sub:transitions} in the Supplementary Text and Supplementary Fig.~\ref{Fig:ORCIDannotations}). From these ordered affiliation trajectories, transitions between fields were identified and recorded. If an individual made two transitions, both were recorded, but transitivity was not used to create additional transitions. That is, a sequence of jobs in $X \to Y \to Z$ would be recorded as only $X \to Y$ and $Y \to Z$, but $X \to Z$ would not be included.  This resulted in a final dataset of 78,798 transitions from 61,108 individuals.

We caution that ORCID users do not constitute a uniform random sample of world scholars~\cite{dasler2017study}. As a result, one might ask whether there are biases in ORCID usership that could invalidate our conclusions. To address this concern, we systematically simulated possible sampling biases in the ORCID data---including biases of the type identified in previous surveys of ORCID membership (e.g., oversampling of STEM fields~\cite{dasler2017study})---to assess the extent to which such sampling biases would affect our conclusions (see Section \ref{section:simulations} in the Supplementary Text). These simulations revealed that the inferences drawn about gender differences in field transitions are valid under a wide range of sampling bias scenarios. Further, there is evidence that field-level gender ratios among ORCID users closely reflect gender ratios reported by other sources. For instance, we found that field-level gender ratios among American PhD recipients, as reported by the U.S. National Science Foundation (NSF)~\cite{NSF2017earned}, were highly correlated ({\it r} = .87) with the gender ratios observed among a subset of ORCID users who approximate the characteristics of NSF's sample (i.e., users with recent affiliations with U.S. universities). These robustness checks suggest that the data and our approach may be used to understand gender differences in career transitions.

\noindent{\bf{Measuring field characteristics: The survey of academics.}}
To test the five explanations under consideration here, we needed to associate academic fields with quantitative measurements of the relevant attributes (e.g., the extent to which they emphasize brilliance). These data were imported from Leslie, Cimpian, Meyer, and Freeland’s recent survey of academics~\cite{leslie2015expectations}. The survey respondents were 1,820 professors, graduate students, and post-doctoral researchers in 30 disciplines including 9 social sciences (e.g., political science, psychology, sociology), 9 humanities (e.g., philosophy, archaeology, art history), and 12 STEM disciplines (e.g., chemistry, computer science,  engineering). The respondents were recruited from 9 geographically diverse universities (5 private, 4 public) from the United States. Participants completed the survey anonymously online and were only asked about their own field (e.g., psychologists were only asked about psychology). The responses from participants within a discipline were averaged. The items included in the survey are listed in Supplementary Table~\ref{Tab:survey}. The correlations between field characteristics are provided in Supplementary Table~\ref{table:corr-mat}.

Questions regarding sampling bias can be raised with respect to the survey dataset as well. Although the survey respondents were not a random sample of U.S. academics, it is likely that their responses nevertheless provide a valid measure of their fields' characteristics. For instance, this measure successfully predicted the proportions of women and African Americans among PhD recipients in the U.S. (as reported by the NSF)~\cite{leslie2015expectations}, a result that has been replicated with other samples of respondents \cite{ito2018factors, meyer2015women} and when adjusting for non-response bias~\cite{berg2010non, leslie2015expectations}. In addition, academics' ratings of their fields' brilliance emphasis (per~\cite{leslie2015expectations}) were highly correlated with a different measure of the same construct---the frequency of the adjectives ``brilliant'' and ``genius'' in 14 million anonymous reviews of instructors in these fields on RateMyProfessors.com \cite{storage2016frequency}. In summary, evidence from the original study \cite{leslie2015expectations} and from subsequent work that built on it \cite{ito2018factors,meyer2015women,storage2016frequency} suggests that the ratings used here capture the characteristics of the fields being rated.   

\noindent{\bf{General analytic strategy.}}
Our goal is to determine what explains gender differences in academics' career trajectories. To pursue this goal, we ask three separate but related questions: \begin{enumerate} 
    \vspace{-2mm}
    \item Which field characteristics explain differences between women and men in their probability of {\it leaving} a field?
    
    \item Which field characteristics explain differences between women and men in their probability of {\it joining} a field?
   
    \item Which field characteristics explain differences between women and men in their {\it transitions across fields}, simultaneously considering the characteristics of the source field and the destination field?
    \vspace{-2mm}
\end{enumerate}

In the next three sections, we describe the results of three models that address the questions above. Throughout, we refer to these models as Main Models I, II, and III, respectively. We then explore two alternative hypotheses that appeal to (i) a field's gender composition and (ii) a field's reliance on mathematics to explain gender differences in academics' career trajectories. Finally, we report a series of analyses that explore the generalizability of our conclusions. In this last set of analyses, we ask the three questions above within key subsets of the data determined by (i) geography, (ii) career stage, and (iii) time.

All analyses were performed in R version 3.3.0 or in Stata 16.1. In all models, standard errors were robust to heteroskedasticity and took into account the clustering in the data, which occurred because some ORCID users made multiple switches. Continuous predictors were mean-centered and scaled by dividing by two standard deviations (SDs;~\cite{gelman2008scaling}). With this scaling, a regression coefficient can be interpreted as indicating the change in the dependent variable that accompanies a change from $-1$ SD to $+1$ SD in the relevant field attribute. 

Because our models simultaneously included multiple field characteristics as predictors, we calculated variance inflation factors (VIFs; Main Model I) or generalized variance inflation factors~\cite{fox1992generalized} (GVIFs; Main Models II and III) to assess multicollinearity. A general rule of thumb is that VIFs $\leq 10$ are acceptable \cite{miles_vif}, although the impact of multicollinearity on estimation accuracy and Type II errors is considerably reduced in large datasets such as ours \cite{mason1991collinearity}. Across analyses, the vast majority of VIFs or GVIFs were below 10, and all were below 20. Given the size of the ORCID dataset, none of these values are reason for concern. For comparison with these models (in which the field characteristics were entered simultaneously), Supplementary Fig.~\ref{Fig:single_var} shows the coefficients from models in which each variable was the sole predictor.

\begin{figure*}[t]
	\includegraphics[width=1.0\linewidth]{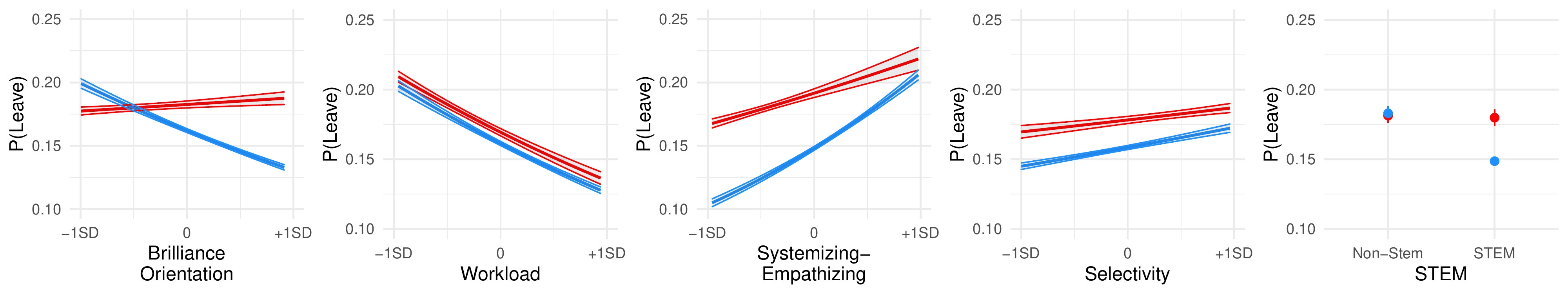}
	\caption{The predicted probability that an academic {\it leaves} a field as a function of five field characteristics (Main Model I), separately for women (red) vs. men (blue). The error bands represent 95\% CIs.\vspace{0.5cm}}
	\label{Fig:stayleave}
\end{figure*}

\section*{Results}

\noindent{\bf{(i) Modeling gender differences in leaving a field.}}
Which field characteristics explain differences between women and men in their probability of leaving a field? To answer this question, we used a logistic regression model to assess the probability that two consecutive affiliations of an ORCID user (e.g., bachelor's degree $\to$ PhD) are in the same field (0 = stay) or in different fields (1 = leave) on the basis of the ORCID user's gender, the five field characteristics, and the two-way interactions between user gender and field characteristics. These interactions provide the answer to our question, since they reveal whether the relationship between a characteristic of a field and the probability that an academic leaves that field differs for women vs. men (see Fig.~\ref{Fig:stayleave} and Main Model I in Supplementary Table~\ref{table:stayleave}).

The results provided support for two of the five hypotheses: Even when adjusting for the other field characteristics, women were more likely than men to leave fields that emphasize brilliance (odds ratio [OR] $=\:1.78\:[1.68, 1.89], p < .00001$) and fields that are in STEM (OR $= 1.27\:[1.17, 1.38], p < .00001$; see Supplementary Table~\ref{table:stayleave}). 

The relation of on-campus workload with the probability of leaving a field did not differ for women and men (OR $= 1.04\:[0.97, 1.11], p = .29$). The relations of a field's emphasis on systemizing vs. empathizing (OR $= 0.62\:[0.57, 0.68], p < .00001$) and selectivity (OR $= 0.91\:[0.87, 0.96], p = .001$) with the probability of leaving did show gender differences, but in the {\it opposite} direction to that hypothesized: Relative to men, women were less---not more---likely to leave fields as the levels of these characteristics increased (see Fig.~\ref{Fig:stayleave} and Supplementary Table~\ref{table:stayleave}).

\begin{figure*}[t]
	\includegraphics[width=1.0\linewidth]{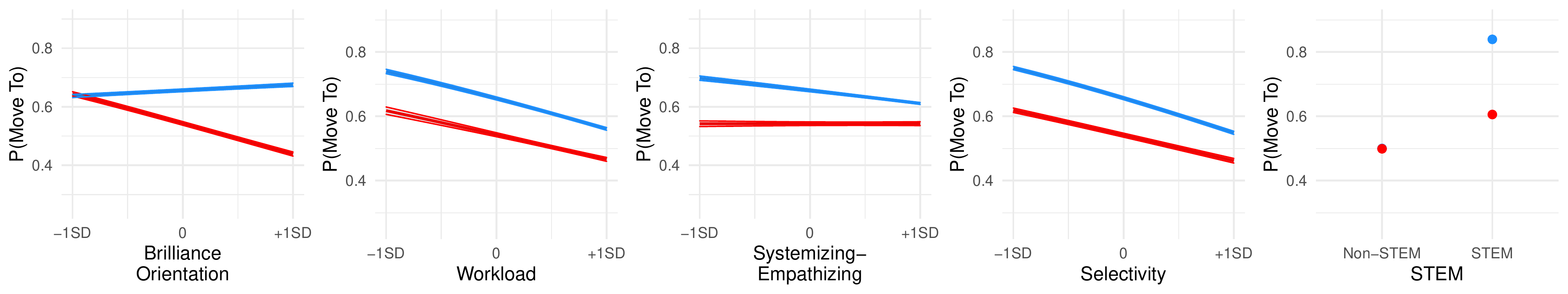}
	\caption{The predicted probability that an academic {\it joins} a field as a function of five field characteristics (Main Model II), separately for women (red) vs. men (blue). As is the default for conditional logistic regression, this probability was calculated assuming a fixed effect (i.e., intercept) of 0. Given that our continuous predictors were mean-centered (i.e., mean = 0), the average predicted probability was $\approx$ .50. The error bands represent 95\% CIs.\vspace{0.5cm}}
	\label{Fig:join}
\end{figure*}

\noindent{\bf{(ii) Modeling gender differences in joining a field.}}
Once academics leave a field, where do they go? Specifically, we asked which field characteristics explain differences between women and men in their probability of joining another field in our sample. To answer this question, we performed a conditional logistic regression, a common means of modeling choices~\cite{wooldridge2010econometric}---in this case, the choice between the 29 possible destination fields in our dataset (excluding the field being departed). Academics' choice to join a particular field was predicted on the basis of their gender, the five field characteristics, and the two-way interactions between academics' gender and field characteristics. 

The results supported the same two hypotheses as above (see Fig.~\ref{Fig:join} and Main Model II in Supplementary Table~\ref{table:choice}): Even when adjusting for all other field characteristics, women were less likely than men to join fields that emphasize brilliance (OR $=\:0.36\:[0.35, 0.38], p < .00001$) and fields that are in STEM (OR $= 0.29\:[0.27, 0.32], p < .00001$). 

The results did not provide support for the other three hypotheses: Relative to men, women were more---not less---likely to join fields that demanded longer working hours (OR $=\:1.20\:[1.12, 1.29], p < .00001$), fields that placed more emphasis on systemizing (vs. empathizing; OR $=\:1.47\:[1.39, 1.55], p < .00001$), and fields that were more selective (OR $=\:1.31\:[1.26, 1.36], p < .00001$; see Fig.~\ref{Fig:join} and Supplementary Table~\ref{table:choice}). It is noteworthy that these gender differences were also considerably smaller in magnitude than those involving the brilliance orientation and STEM variables. For example, the odds ratio for the interaction between gender and brilliance orientation was approximately twice as high as the odds ratios for the interactions with workload, systemizing--empathizing, and selectivity. 

\begin{figure*}[t]
	\includegraphics[width=1.0\linewidth]{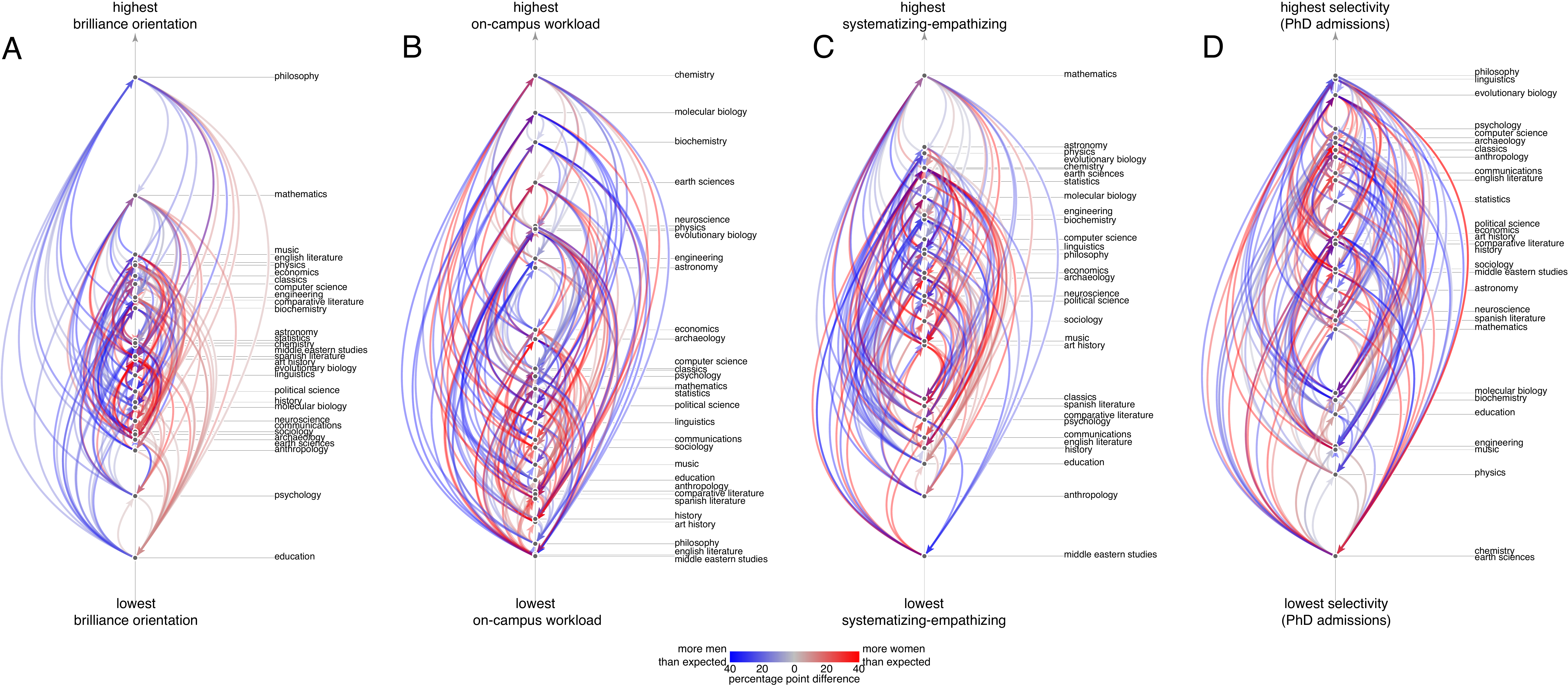}
	\caption{Transitions between fields as a function of the fields’ brilliance orientation (A), on-campus workload (B), systemizing--empathizing focus (C), and selectivity (D). The more red vs. blue an arc, the greater the proportion of women vs. men, respectively, transitioning between the two fields relative to expectations; all plots share the same color mapping. For each attribute (A-D), fields are displayed vertically in descending order. Transitions up and down the gradient are displayed on the left and right of each vertical axis, respectively. For simplicity, we only display the arcs for which the gender of the transitioning individuals differed significantly from what would be expected given the gender composition of the source field, according to a $z$-test for differences in proportions. The reliance of the fifth hypothesis on a dichotomous STEM vs. non-STEM distinction made it less amenable to inclusion in this figure.}
	\label{Fig:spaghetti}
\end{figure*}

\noindent{\bf{(iii) Modeling gender differences in field transitions.}}
Examining which fields women and men are differentially likely to leave without also considering where they go (as in Main Model I) is underinformative, and so is examining which fields women and men are differentially likely to join without also considering where they came from (as in Main Model II). For instance, if an academic leaves a field with a certain brilliance-orientation score, it is important for our purposes to take into account whether they switch to a field that is higher or lower in its brilliance orientation. Analogously, we would draw different conclusions if an academic who joined a field with a certain brilliance-orientation score came from a field that was higher vs. lower in its brilliance orientation than the destination field. The third and final model (Main Model III) addresses this shortcoming of the first two models by modeling gender differences in {\it field transitions}, simultaneously considering the characteristics of the source field and the destination field.

Fig.~\ref{Fig:spaghetti} depicts the field transitions observed among the academics in our dataset. Looking at Panel A (brilliance orientation), we see that most upstream (low $\to$ high) arcs are blue in color and most downstream (high $\to$ low) arcs are red in color. This indicates that men are more likely than women to move up the brilliance-orientation gradient (i.e., toward fields that place greater emphasis on this characteristic) and, conversely, that women are more likely than men to move down this gradient. A similar analysis applies to Panel C (systemizing--empathizing), whereas for Panels B (workload) and D (selectivity) the downstream and upstream transitions appear more gender-balanced. 

\begin{figure*}[t]
	\includegraphics[width=1.0\linewidth]{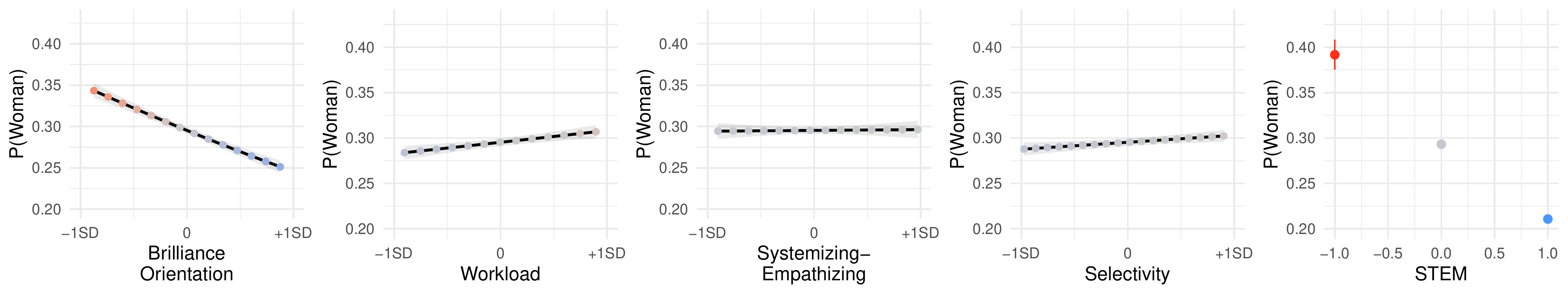}
	\caption{The predicted probability that an academic transitioning between two fields is a woman as a function of the gradients for five field characteristics (Main Model III). Gradients are the differences between the destination field's value on each characteristic and the source field's value. The more red vs. blue a dot, the higher vs. lower, respectively, the probability that the transitioning individual is a woman. The error bands represent 95\% CIs.}
	\label{Fig:flow}
\end{figure*}

For a quantitative test of the five explanations for gender differences in field transitions, we used a logistic regression in which we predicted the gender of an individual moving between two fields on the basis of gradients for the five field characteristics of interest. Each gradient was calculated as the difference between the value of a characteristic (e.g., brilliance orientation) for the destination field and the value of that same characteristic for the source field. A positive gradient therefore means that an individual is transitioning to a field that displays more of a certain characteristic than the field the individual came from. As a result, a positive regression coefficient for a particular gradient signifies that women are more likely to transition into fields with higher values of that characteristic (i.e., upstream), and a negative coefficient signifies downstream movement for women. The model also included indicator variables for all but one of the 30 source fields. This analytic strategy adjusts for differences in the gender composition of the source fields and thus for differences among them in the probability that the individuals who switch are women. Finally, we note that although we describe the results of this model as reflecting the odds of a transitioning individual being a woman (vs. a man), the results obviously reflect men’s field transitions as much as they do women’s. Our descriptive focus on women should not be interpreted as a substantive claim that gender segregation is solely a function of women’s career decisions~\cite{miller1991gender,cheryan2017some}.

Consistent with the previous two models, the results indicated that women were significantly more likely than men to transition toward fields that were lower in their brilliance orientation (OR $=\:0.60\:[0.56, 0.65], p < .00001$) and toward non-STEM fields (OR $=\:0.64\:[0.60, 0.69], p < .00001$; see Fig.~\ref{Fig:flow} and Main Model III in Supplementary Table~\ref{table:flow}). The results for the remaining gradients did not support their respective hypotheses: The systemizing--empathizing gradient did not predict the gender of academics switching between fields (OR $=\:1.01\:[0.92, 1.11], p = .85$), while the coefficients for workload (OR $=\:1.13\:[1.06, 1.21], p = .0002$) and selectivity (OR $=\:1.08\:[1.02, 1.14], p = .009$) suggested that relative to men, women were more likely to move {\it up} these gradients---toward fields that have higher workloads and are more selective. These relationships were relatively modest in magnitude (see Fig.~\ref{Fig:flow}). For instance, the odds ratio for the brilliance orientation gradient was approximately 50\% higher than the odds ratios for the workload and selectivity gradients.

\noindent{\bf{Alternative hypothesis I: Homophily.}}
Next, we investigated the homophily principle---that is, the tendency to gravitate toward fields with more people of one's gender~\cite{mcpherson2001birds}---as an alternative explanation for our results. Homophily is a conservative standard of comparison, in that the processes under consideration here (e.g., women moving toward fields that are lower in brilliance orientation) result in homophily themselves, so the gender differences in career trajectories explained by homophily could be due in part to these other processes. 

To test this alternative explanation, we added a variable tracking the gender composition of each field (calculated from the ORCID data\footnote{The homophily variable was computed using a much larger dataset than the dataset used to analyze field transitions--a dataset that also included the
academics who never transitioned out of their field (550,961 researchers) rather than just those
who transitioned between fields (61,108 researchers).}) to the three models above. Specifically, for the models examining the probability of leaving (Main Model I) or joining (Main Model II) a field, we added a variable consisting of the log odds of being a woman in the fields being left or joined, respectively, as well as this variable's interaction with the gender of the ORCID user. For the model on field transitions (Main Model III), we added a gradient for homophily, calculated as the difference between the log odds of being a woman in the destination field and the analogous log odds for the source field. 

As expected, homophily was a significant predictor of academics' career trajectories: The more women there were in a field, the less likely women were to {\it leave} that field relative to men (OR $=\:0.67\:[0.62, 0.73], p < .00001$) and the more likely they were to {\it join} it (OR $=\:2.18\:[2.07, 2.31], p < .00001$). Similarly, women moved up the homophily gradient, toward fields with more women (OR $=\:1.97\:[1.82, 2.13], p < .00001$; see Model B in Supplementary Tables~\ref{table:stayleave},~\ref{table:choice}, and~\ref{table:flow}).

Of the five variables of primary interest, only brilliance orientation remained significant in all three models after we included the homophily variable: The more a field emphasized brilliance, the more likely women were to leave it, relative to men, even after accounting for homophily (OR $=\:1.27\:[1.19, 1.36], p < .00001$) and the less likely they were to join it (OR $=\:0.53\:[0.50, 0.55], p < .00001$). Also, as in Main Model III (Supplementary Table~\ref{table:flow}), women were more likely than men to move down the brilliance orientation gradient (OR $=\:0.90\:[0.82, 0.99], p = 0.022$). 

\noindent{\bf{Alternative hypothesis II: Math-intensiveness.}}
So far, we have found that fields with stronger emphasis on raw intellectual talent (``brilliance'') tend to lose women and gain men across successive career transitions---a mechanism that contributes to the patterns of gender segregation observed in academia. However, an alternative interpretation for this result is that beliefs about the importance of brilliance in a field are simply a symptom of the extent to which success in that field depends on mathematical ability~\cite{ginther2015comment,cimpian2015response}. Because STEM and non-STEM fields also differ in the extent to which they rely on mathematics, the math-intensiveness of a field may also explain the observed differences between STEM and non-STEM fields in their ability to recruit and retain women vs. men.

To test this alternative explanation, we added a variable corresponding to it in Main Models I, II, and III: namely, the average Quantitative GRE scores of graduate applicants to each field (as reported by the Educational Testing Service), which can serve as a proxy for the field's emphasis on mathematics~\cite{ginther2015comment,cimpian2015response}.  Specifically, for the models examining the probability of leaving (Main Model I) or joining (Main Model II) a field, we added the scores of applicants to the fields being left or joined, respectively, as well as this variable’s interaction with the gender of the ORCID user. For the model on field transitions (Main Model III), we added a gradient variable, calculated as the difference between the average Quantitative GRE scores of applicants to the destination field and the analogous average for the source field. These variables were mean-centered and scaled by dividing by 2 SDs~\cite{gelman2008scaling}, like the other continuous field attributes, to facilitate comparison of effect sizes. We used the Quantitative GRE data from references~\cite{ginther2015comment} and~\cite{cimpian2015response}; GRE scores were available for all fields except two: linguistics and music theory and composition.

Consistent with the alternative hypothesis, the higher a field's Quantitative GRE score, the more likely women were to leave that field relative to men (OR $=\:1.19\:[1.09, 1.31], p = .0002$) and the less likely they were to join it (OR $=\:0.71\:[0.67, 0.76], p < .00001$). Similarly, women were more likely than men to move down the Quantitative GRE gradient, toward less math-intensive fields (OR $=\:0.71\:[0.65, 0.77], p < .00001$; see Model C in Supplementary Tables~\ref{table:stayleave},~\ref{table:choice}, and~\ref{table:flow}).

However, contrary to this alternative hypothesis, both brilliance orientation and STEM explained gender differences in career trajectories (in the hypothesized direction) even after partialing out the variance attributable to the Quantitative GRE, in all three models. In terms of effect sizes, the relevant odds ratios were 2\% to 73\% higher for brilliance orientation than for the Quantitative GRE across the three models, and 6\% lower to 81\% higher for STEM than for the Quantitative GRE.

\begin{figure*}
	\includegraphics[width=0.82\textwidth]{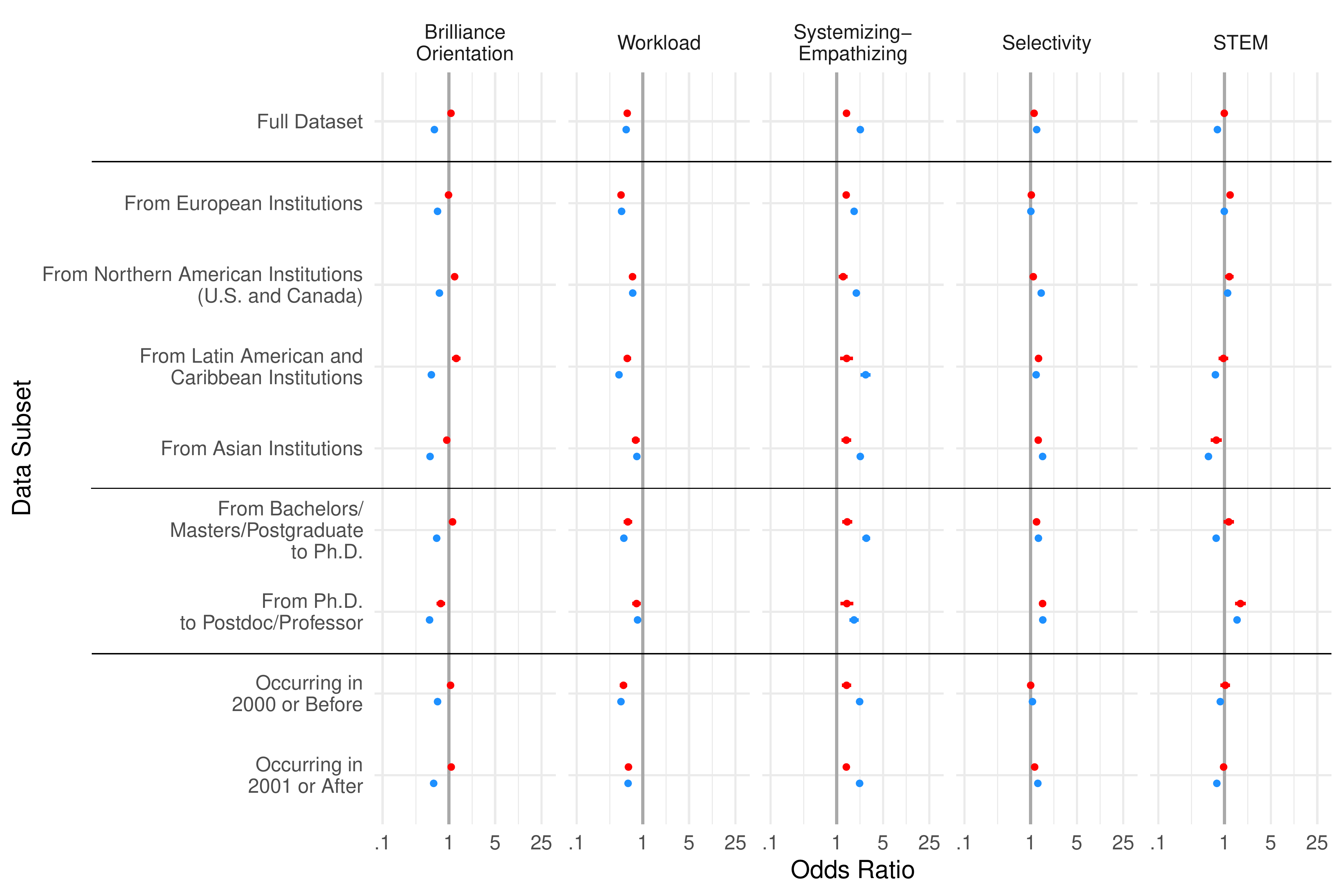}
	\caption{Odds ratios for women (red) and men (blue) from {\bf Main Model I} (``who leaves?'') in the full dataset (see top row) and across key subsets of the data. Each row corresponds to the results from Main Model I applied to a different subset of the data. Error bars represent 95\% CIs. The $x$ axis is on a logarithmic scale.}
	\label{Fig:gen:stay-leave}
\end{figure*}

\begin{figure*}
	\includegraphics[width=0.82\textwidth]{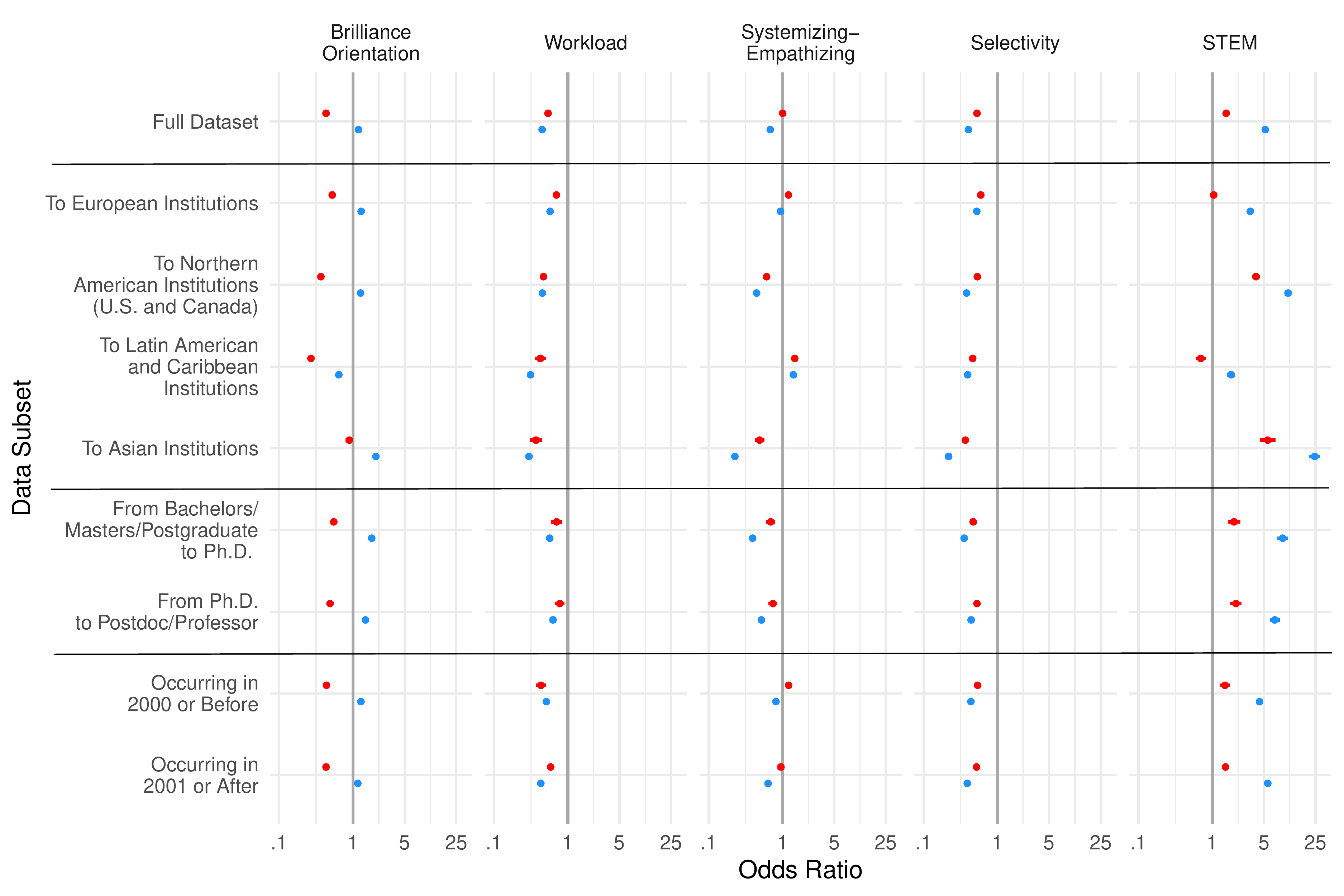}
	\caption{Odds ratios for women (red) and men (blue) from {\bf Main Model II} (``who joins?'') in the full dataset (see top row) and across key subsets of the data. Each row corresponds to the results from Main Model II applied to a different subset of the data. Error bars represent 95\% CIs. The $x$ axis is on a logarithmic scale.}
	\label{Fig:gen:choice}
\end{figure*}

\begin{figure*}
	\includegraphics[width=0.82\textwidth]{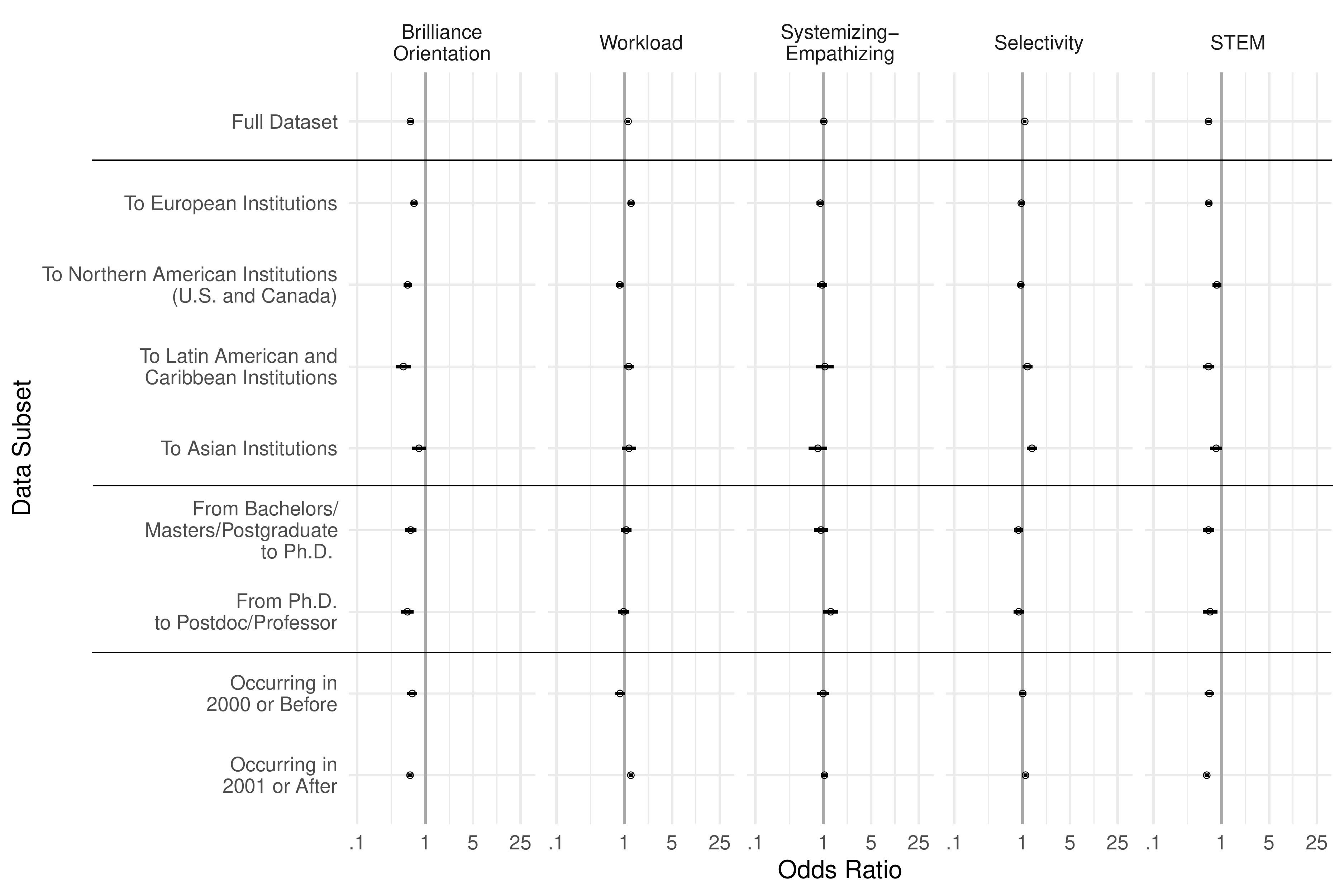}
	\caption{Odds ratios from {\bf Main Model III} (field transitions) in the full dataset (see top row) and across key subsets of the data. If a gradient has an OR below 1, that indicates that women are more likely than men to move down that gradient (and vice-versa). Each row corresponds to the results from Main Model III applied to a different subset of the data. Error bars represent 95\% CIs. The $x$ axis is on a logarithmic scale.}
	\label{Fig:gen:flows}
\end{figure*}

\noindent{\bf{Tests of generalizability: Geography, career stage, and time.}}
In the last set of analyses, we explored whether the results above---in particular, the differences between women’s and men’s career trajectories as a function of fields' brilliance orientation and STEM status---generalize with respect to geography, career stage, and time. We did so by applying Main Models I, II, and III to key subsets of the data: (i) career transitions involving institutions from Europe, Northern America (U.S. and Canada), Latin America and the Caribbean, and Asia, (ii) career transitions from bachelor’s or master’s programs to PhD programs and from PhD programs to postdoctoral positions or professorships, and (iii) career transitions that occurred before (and including) the year 2000 and after 2000. (Using other years as split points led to similar results.) The results are illustrated in Figs.~\ref{Fig:gen:stay-leave},~\ref{Fig:gen:choice}, and~\ref{Fig:gen:flows} for Main Models I, II, and III, respectively.

Fig.~\ref{Fig:gen:stay-leave} shows the results for Main Model I (``who leaves?'') and reveals that fields' brilliance orientation  consistently predicted greater odds of leaving for women (the red ORs) relative to men (the blue ORs). This was the case across every subset examined: by geography (rows 2--5), career stage (rows 6 and 7), and time (rows 8 and 9). STEM was the only other variable that consistently showed the hypothesized gender differences, but the OR differences were noticeably smaller than those for brilliance orientation. The other variables either showed no consistent difference between women's and men's ORs (workload and selectivity) or somewhat consistent differences contrary to the hypothesized direction (systemizing--empathizing).

Fig.~\ref{Fig:gen:choice} shows the results for Main Model II (``who joins?'') and reveals that brilliance orientation and STEM were the only variables that showed the predicted gender differences---with greater odds of joining brilliance-oriented and STEM fields for men relative to women---across all subsets of the data. This time, the gender differences in ORs were larger for STEM than for brilliance orientation. The other three variables showed small differences in the unpredicted direction across most subsets.

Finally, Fig.~\ref{Fig:gen:flows}, which plots the results for Main Model III (field transitions), reveals a pattern of results similar to those above: The only gradients whose ORs consistently differed from 1 across subsets (indicating gender differences) were those for brilliance orientation and STEM.

These analyses suggest that a field's emphasis on brilliance and STEM status are---and have been---sources of gender segregation in academia, predicting higher attrition and lower recruitment rates for women at all career stages and across the globe.

\section*{Discussion}

\noindent{\bf{Limitations and future directions.}}
An important limitation of this work is that the field characteristics were measured from a sample of U.S. academics~\cite{leslie2015expectations} rather than from a global sample. If these characteristics varied across countries or regions, we would not be able to take these variations into account when analyzing the patterns underlying gender segregation. In light of this limitation, it is striking that fields' brilliance orientation nevertheless emerged as a reliable predictor of career transitions. This measure asks whether success in a field depends on possessing qualities such as a ``special aptitude'' or an ``innate gift or talent,'' which are subjective, ill-defined judgments that one might reasonably expect to vary cross-culturally. It is also noteworthy that the variables that did not reliably explain the observed gender differences in career transitions (i.e., workload, systemizing--empathizing, and selectivity) did not do so even among the academics from the U.S. and Canada (see Figs.~\ref{Fig:gen:stay-leave},~\ref{Fig:gen:choice}, and~\ref{Fig:gen:flows})---a sample that is culturally similar to that which rated the field characteristics. Thus, the lack of evidential support for these explanations cannot be wholly attributable to issues of cross-cultural validity. Nevertheless, future surveys of academics with a broader geographical scope could, when combined with ORCID profiles, provide a more precise test of the explanations considered here. 

Another limitation of the data is that the field attributes were measured at a single time point. Notably, there was no drop in the predictive validity of the field characteristics, as reported in 2015~\cite{leslie2015expectations}, for transitions that occurred 15 years or more earlier (compare the bottom two rows of Figs.~\ref{Fig:gen:stay-leave},~\ref{Fig:gen:choice}, and~\ref{Fig:gen:flows}). Nevertheless, the characteristics of a field may change over time~\cite{cheryan2017some}, so research that measures field characteristics at multiple time points and relates them dynamically to the observed levels of gender segregation across fields would be valuable.

Although we considered STEM disciplines as a group, there are important differences among this group in the extent of gender segregation and the climate that women face~\cite{cheryan2017some,national2018sexual,cimpian2020understanding}. In particular, computer science, engineering, and physics remain more segregated than the rest of the STEM fields and are particularly likely to exhibit ``masculine cultures'' that undermine women's psychological safety~\cite{cheryan2017some}. To explore this finer-grained distinction with our data, we redefined the STEM variable to have a narrower scope (1 = astronomy, computer science, engineering, or physics; 0 = all other fields) and re-ran Main Models I, II, and III. The results suggested that women are {\it not} more likely to leave these four fields (vs. the others) than men are (OR $=\:1.01\:[0.96, 1.06], p = .77$; Main Model I), but they are in fact less likely to join them (OR $=\:0.67\:[0.64, 0.71], p < .00001$; Main Model II). This combination of results is consistent with recent evidence that these fields have more of a recruitment than a retention problem~\cite{cheryan2017some,cimpian2020understanding}. In the model on field transitions (Main Model III), women were significantly more likely than men to transition toward fields other than these four (OR $=\:0.66\:[0.63, 0.69], p < .00001$). (We note that the corresponding coefficients for the brilliance orientation variable remained significant in all three models, $p\text{s} < .00001$.) There are, of course, many more interesting questions to ask on this topic. In future work, it will be important to use the ORCID dataset to examine in greater detail the differences among STEM fields with an eye toward understanding why some have moved toward gender desegregation while others have not.

There may also be heterogeneity in the mechanisms underlying gender segregation. Although the dimensions along which we split the data here did not reveal substantial levels of heterogeneity, we did see some hints of it. For instance, Figs.~\ref{Fig:gen:choice} and~\ref{Fig:gen:flows} suggest a subtle shift in the effect of a field's workload across time: In earlier, pre-2000 field switches, women were less likely than men to move into fields with high workloads (holding all other attributes constant), whereas in more recent transitions, this difference reversed---a trend that is consistent with other changes observed over the last few decades in gender roles and attitudes~\cite{eagly2019gender,Blau2018WomenWorkFamily}. This example aside, our present focus on broad, cross-cutting explanations for gender segregation may have overlooked some of the heterogeneity in this global longitudinal dataset. By making our code and data available to other researchers, we hope to facilitate work that delves deeper into heterogeneity and its sources, as well as work that brings additional explanations (beyond those considered here) to bear on this rich dataset.

\noindent{\bf{Summary and implications.}}
Gender segregation in academia is a persistent, global issue. However, much of the research on this topic has been narrower in scope than the phenomenon it set out to explain. Using the single largest dataset of academic profiles, supplemented with ratings from a recent survey of academics, we investigated the differential migration of women and men between fields across a range of fields in STEM, the social sciences, and the humanities.

Our results suggest that a field’s gender composition may be explained in part by the extent to which it values brilliance. Greater emphasis on raw intellectual talent predicted greater numbers of women leaving a field and greater numbers of men joining it. Although there is no compelling evidence that women and men actually differ on this trait~\cite{charlesworth2019gender}, common stereotypes nevertheless associate brilliance and genius with men more than women~\cite{bian2017gender,storage2020adults}. These stereotypes might lead women to opt out of fields where brilliance is valued~\cite{bian2018messages}, and they also might prompt some members of these fields to doubt women’s ability to succeed, depriving them of opportunities for advancement~\cite{bian2018evidence, moss2012science}. In these ways, a belief that on the surface seems unbiased---namely, the belief that success requires brilliance---may have a differential impact on women’s and men’s career trajectories and, in turn, may exacerbate segregation in the fields where this belief is widely endorsed. The relationship between the relative endorsement of this belief across fields and women’s disproportionate departures (as well as men’s disproportionate joining) was strikingly robust across geography, career stages, and time. 

We also found that women were more likely to migrate out of STEM fields and men were more likely to migrate into them, even when adjusting for other field attributes such as brilliance orientation, emphasis on systemizing vs. empathizing, and reliance on mathematics. This finding is consistent with arguments of a ``leaky pipeline'' for women in STEM~\cite{Alper1993pipeline} and puts recent reports that STEM fields (in the U.S.) no longer lose more women than men in a broader perspective~\cite{ceci2014women,miller2015bachelor}. It remains to be determined what attributes of STEM fields explain the gender-differentiated transitions out of and into them. Candidates include, among others, the masculine culture of some of these fields~\cite{berdahl2018MCC,cheryan2017some} and the elevated levels of sexual harassment directed at women in some STEM fields, such as engineering~\cite{national2018sexual}.

From a policy standpoint, the present findings suggest that intervention efforts might fruitfully be targeted at the belief that raw intellectual talent is required for success in a field. Although women and men do not differ in their intellectual potential, cultural stereotypes suggest that they do, which makes the environment of brilliance- and talent-oriented fields unwelcoming for many capable young women. By redirecting the messages being sent to young people away from a focus on raw, untutored talent and toward the concrete skills they will need to be successful~\cite{dweck2008mindset, yeager2019national}, many fields may be in a better position to attract and retain a diverse workforce.


\vspace{.5cm}

\noindent{\bf Data availability} \\
\normalsize{All data are available at {\href{https://github.com/kennyjoseph/ORCID_career_flows}{https://github.com/kennyjoseph/\\ORCID\_career\_flows}}.}

\vspace{.25cm}

\noindent{\bf Code availability} \\
\normalsize{All Python, R, and Stata code are available at {\href{https://github.com/kennyjoseph/ORCID_career_flows}{https://github.\\com/kennyjoseph/ORCID\_career\_flows}}.}

\vspace{.5cm}

\noindent{\bf Acknowledgements} \\
The authors are grateful to Joe Cimpian, Aaron Clauset, Fred Feinberg, David Garcia, Jillian Lauer, Melis Muradoglu, Jessica Nordell, Carrie Shandra, Molly Tallberg, Andrea Vial, and K. Hunter Wapman for helpful comments on previous drafts of the manuscript. This work was supported in part by grants from the U.S. National Science Foundation to Andrei Cimpian (BCS-1733897) and Daniel B. Larremore (SMA-1633747), the Russell Sage Foundation to Aniko Hannak (92-17-03), and the Air Force Office of Scientific Research to Daniel B. Larremore (FA9550-19-1-0329). ORCID was not involved in designing, conducting, or writing up this research; data were used with the appropriate permissions and according to ORCID’s terms of service. 

\vspace{.25cm}

\noindent{\bf Author contributions} \\
All authors conceived of the research, conducted the research, and wrote the manuscript. 

\vspace{.25cm}

\noindent{\bf Competing interests}  \\
The authors declare no competing interests.

\clearpage
\setcounter{equation}{0}
\setcounter{figure}{0}
\setcounter{table}{0}
\setcounter{section}{0}
\renewcommand{\theequation}{\arabic{equation}}
\renewcommand{\thefigure}{\arabic{figure}}
\renewcommand{\thetable}{\arabic{table}}
\renewcommand{\thesection}{\arabic{section}}
\renewcommand{\thesubsection}{\hspace{-1mm}-\Alph{subsection}}
\renewcommand{\theHtable}{Supplement.\thetable}
\renewcommand{\theHfigure}{Supplement.\thefigure}

\renewcommand\labelenumi{(\arabic{enumi})}
\renewcommand\theenumi\labelenumi

\renewcommand\tablename{Supplementary Table}
\renewcommand\figurename{Supplementary Fig.}

\titleformat{\section}[block]{\bfseries\filcenter\large}{\thesection\hspace{2pt}.}{4pt}{}

\titleformat{\subsection}[block]{\bfseries\filcenter}{\thesection\hspace{2pt}\thesubsection.}{4pt}{}

\onecolumngrid

\section*{\huge{Supplementary Information}}
\section*{\Large{Explaining Gender Differences in Academics' Career Trajectories}}
\section*{Supplementary Tables and Figures Referenced in the Main Text}
\setcounter{page}{1}

\vspace{2cm}

\begin{figure*}[h]
	\includegraphics[width=1.0\linewidth]{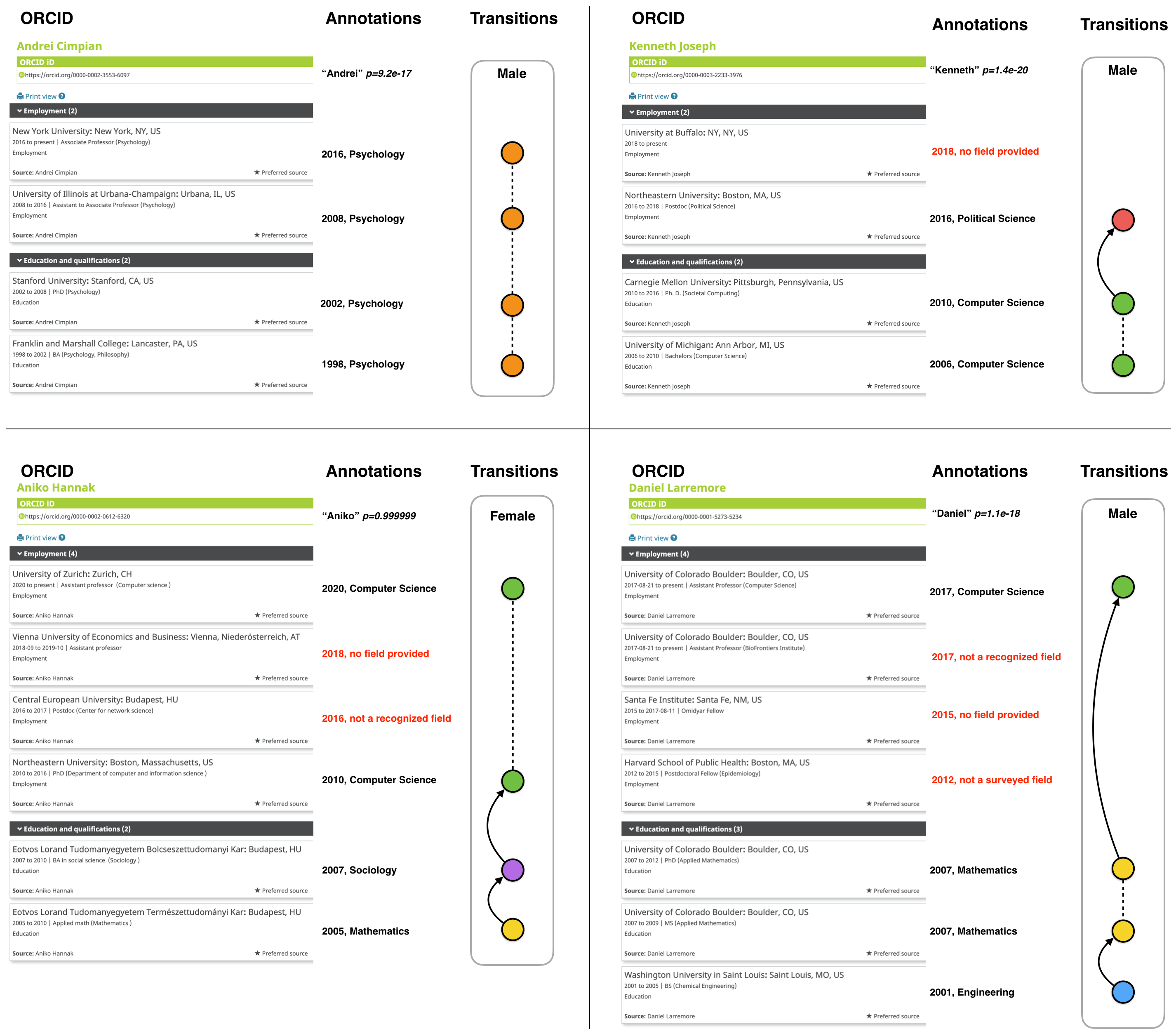}
	\caption{Public ORCID profiles for the authors of the study, with annotations illustrating choices made during data processing.}
	\label{Fig:ORCIDannotations}
\end{figure*}

\begin{table*}[h]
\caption{The items from Leslie, Cimpian, Meyer, and Freeland’s survey of academics~\cite{leslie2015expectations}.}
\label{Tab:survey}
\footnotesize
\begin{tabular}{ p{0.5cm} p{0.5cm} l }
\hline
\vspace{-6pt} \\
\multicolumn{3}{l}{{\bf Field-specific Ability Beliefs (Brilliance Orientation)}$^a$}                                                                                          \\
                         & \multicolumn{2}{l}{Being a top scholar of [discipline] requires a special aptitude that just can’t be taught.}                            \\
                         & \multicolumn{2}{l}{If you want to succeed in [discipline], hard work alone just won’t cut it; you need to have an innate gift or talent.} \\
                         & \multicolumn{2}{l}{With the right amount of effort and dedication, anyone can become a top scholar in [discipline]. (R)}                                                                                                                         \\
                         & \multicolumn{2}{l}{When it comes to [discipline], the most important factors for success are motivation and sustained effort; raw ability is secondary. (R)}                                                                                                                         \\
\multicolumn{3}{l}{{\bf Workload}$^b$}                                                                                                                                         \\
                         & \multicolumn{2}{l}{Approximately how many hours a week do you spend working:}                                                                                                                        \\
                         &                                                                       & In your office, lab, classroom, or otherwise on campus?                                                                     \\
                         &                                                                       & Off campus (e.g., home, coffee shop, other remote site)?                                                                     \\
\multicolumn{3}{l}{{\bf Selectivity}$^c$}                                                                                                                                      \\
                         & \multicolumn{2}{l}{Roughly what percentage of applicants are accepted into your department’s PhD program in a typical year? (R)}                                                                                                                          \\
\multicolumn{3}{l}{{\bf Systemizing vs. Empathizing}$^d$}                                                                                                                      \\
                         & \multicolumn{2}{l}{Please rate the extent to which the following processes are involved in doing scholarly work in [discipline]:}                                                                                                                         \\
                         &                                                                       & Identifying the abstract principles, structures, or rules that underlie the relevant subject matter (Systemizing)                                                                     \\
                         &                                                                       & Analyzing the relevant subject matter and constructing a systematic understanding of it (Systemizing)                                                                     \\
                         &                                                                       & Having a refined understanding of human thoughts and feelings (Empathizing)                                                                     \\
                         &                                                                       & Recognizing and responding appropriately to people’s mental states (Empathizing)                                                                     \\
\vspace{-6pt} \\
\hline
\multicolumn{3}{l}{\footnotesize {\it Note.} (R) indicates items that were reverse scored.}                                                                                                                                                    \\
\multicolumn{3}{l}{\footnotesize $^a$Responses to these items were given on a 7-point scale (1 = strongly disagree to 7 = strongly agree).}                                                                                                                                                    \\
\multicolumn{3}{l}{\footnotesize \begin{tabular}[c]{@{}l@{}} $^b$Responses to these items were given on an 8-point scale (1 to 8, 1-7 corresponding to 10-hour increments, and 8 corresponding to $>$70 hours). \\
\quad   Because the {\it off}-campus hours variable did not predict gender gaps in prior work~\cite{leslie2015expectations}, we focused on the {\it on}-campus hours variable in our analyses.\end{tabular}}                                                                                                                                                    \\
\multicolumn{3}{l}{\footnotesize \begin{tabular}[c]{@{}l@{}} $^c$Responses to these items were given on a 10-point scale (1 to 10, each number corresponding to a 10\% increment). There were two additional\\ 
\quad   options for ``don’t know'' and ``no PhD program''. This item was administered only to faculty respondents.\end{tabular}}                                                                                                                                                    \\
\multicolumn{3}{l}{\footnotesize \begin{tabular}[c]{@{}l@{}} $^d$Responses to these items were given on a 7-point scale (1 = never involved to 7 = highly involved). Each respondent’s empathizing ratings were\\
\quad averaged, and this average was then subtracted from the average of their systemizing ratings. Finally, the difference scores of all respondents\\
\quad from a certain field were averaged into that field’s systemizing--empathizing score, with higher values indicating more emphasis on systemizing\\
\quad relative to empathizing.\end{tabular}}
                                                       
\end{tabular}
\end{table*}

\def\arraystretch{1.25}
\begin{table*}[h]
    \captionsetup{justification=centering}
  \caption{Correlations among the five field characteristics.} 
  \label{table:corr-mat} 
\begin{tabular}{p{4.5cm}SSSSSp{0.05cm}} 
 \hline \hline
Variable & 
{\centering\text{1.}} & 
{\centering\text{2.}} & 
{\centering\text{3.}} & 
{\centering\text{4.}} & 
{\centering\text{5.}} &\\
\hline

1. Brilliance Orientation & 1.00 \\
 
2. Workload & -.09 & 1.00\\ 

3. Systemizing--Empathizing & .34$^{\sim}$ & .70$^{***}$ & 1.00\\ 

4. Selectivity & .07 & -.51$^{**}$ & -.30 & 1.00\\ 

5. STEM  & .14 & .79$^{***}$ & .79$^{***}$ & -.46$^{*}$ & 1.00\\

\hline \hline 
\multicolumn{6}{l}{{\it N} = 30 fields. $^{\sim}{\it p} < 0.10$; $^{*}{\it p} < 0.05$; $^{**}{\it p} < 0.01$; $^{***}{\it p} < 0.001$}   \\ 
\end{tabular} 
\end{table*}

\begin{figure*}[t]
	\includegraphics{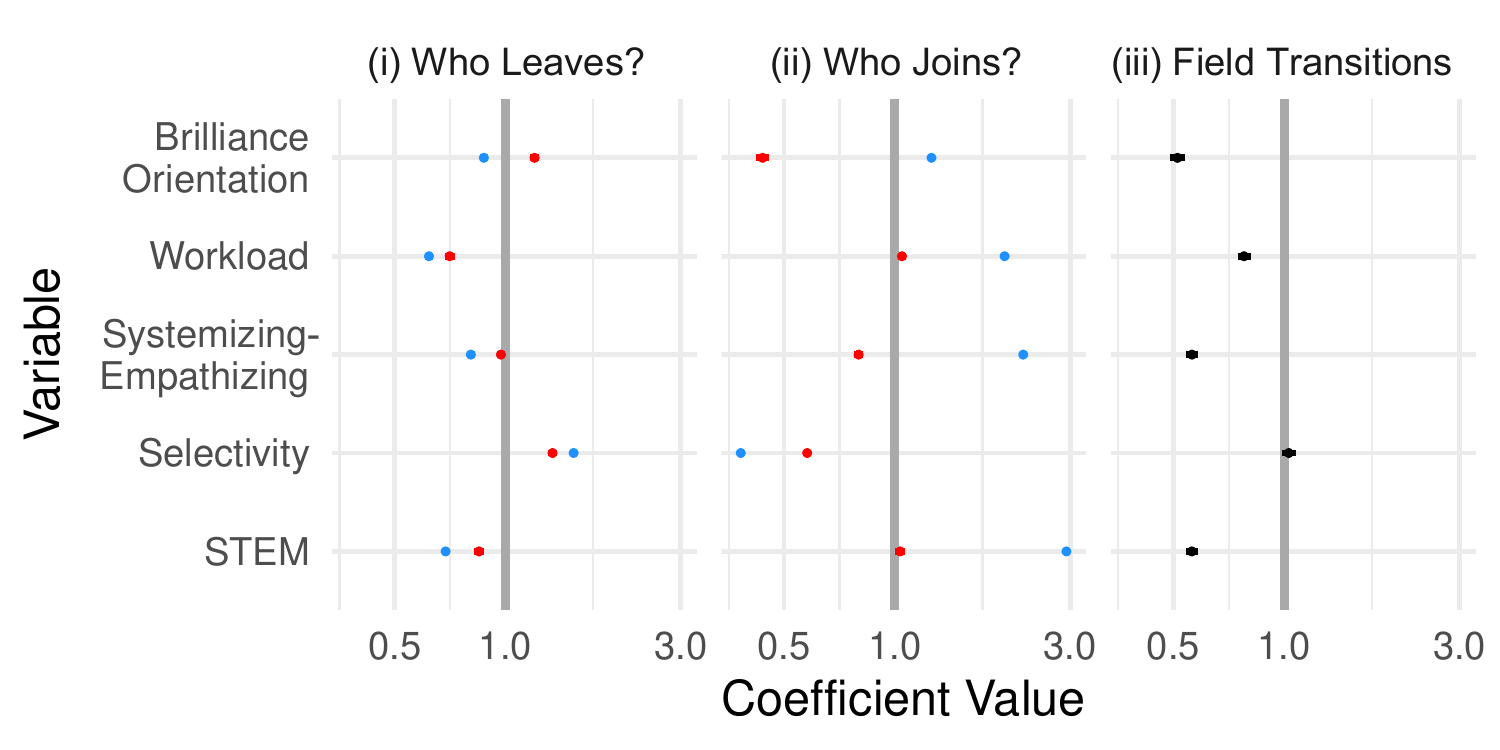}
	\caption{Coefficients (expressed as ORs) from models in which each of the field characteristics was entered as a solo predictor. The dots in panels (i) and (ii) represent the relationship of the relevant field characteristic with the probability of leaving and joining a field, respectively, for women (red) and men (blue). For instance, the fact that the OR is higher for women than for men for the Brilliance Orientation variable in panel (i) suggests that women are more likely than men to leave a field as its brilliance orientation increases. For panel (iii), if a gradient has an OR below 1, that indicates that women are more likely than men to move downstream for that attribute. If a gradient has an OR above 1, that indicates that women are more likely than men to move upstream for that attribute. For instance, the fact that the OR for the Brilliance Orientation variable in panel (iii) is below 1 suggests that women are more likely than men to move down the brilliance orientation gradient. Error bars represent 95\% CIs (but are not easily visible because the size of the dataset allows precise estimation). The $x$ axis is on a logarithmic scale.}
	\label{Fig:single_var}
\end{figure*}


\begin{table*}
  \caption{Coefficients (and cluster--robust standard errors) from logistic regressions that model the probability that an ORCID user {\bf leaves a field} based on their gender (0 = man, 1 = woman) and the five field characteristics of primary interest (Model A), as well as homophily (Model B) or Quantitative GRE scores (Model C).} 
  \label{table:stayleave} 
\begin{tabular}{p{5.5cm}Sp{1.25cm}p{0.1cm}Sp{1.25cm}p{0.1cm}Sp{1.25cm}} 
 \hline \hline

& \multicolumn{2}{c}{\enspace \enspace Model A} && \multicolumn{2}{c}{\enspace \enspace Model B} && \multicolumn{2}{c}{\enspace \enspace  Model C} \\
& \multicolumn{2}{c}{\enspace \enspace Main Model I\vspace{1mm}} && 
\multicolumn{2}{c}{\enspace \enspace  Alternative: Homophily} && 
\multicolumn{2}{c}{\enspace \enspace Alternative: Quant GRE}\\
\hline
 
 Brilliance Orientation & -0.506$^{***}$ & ($0.019$) && 0.105$^{***}$ & ($0.022$) && -0.130$^{***}$ & ($0.020$)\\ 
 Workload & -0.580$^{***}$ & ($0.021$) && -0.632$^{***}$  & ($0.019$) && -0.658$^{***}$ & ($0.020$) \\ 
 Systemizing--Empathizing & 0.819$^{***}$ & ($0.027$) && 0.480$^{***}$ & ($0.027$) && 1.230$^{***}$ & ($0.028$) \\ 
 Selectivity & 0.209$^{***}$ & ($0.017$)&& 0.018 & ($0.018$) && 0.115$^{***}$ & ($0.018$)\\ 
 STEM & -0.249$^{***}$ & ($0.022$) && 0.731$^{***}$ & ($0.031$) && 0.360$^{***}$ & ($0.028$) \\ 
 Homophily (log odds woman per field) &&&& 1.248$^{***}$  & ($0.025$) \\ 
 Quantitative GRE &&&&&&& -1.314$^{***}$ & ($0.027$) \\ 
 Is Woman & -0.034 & ($0.028$) && 0.243$^{***}$   & ($0.037$) && 0.007 & ($0.036$)\vspace{0.4cm} \\ 
 Is Woman $\times$ \underline{\hspace{1.5cm}} \\
  \hspace{0.75cm}Brilliance Orientation & 0.577$^{***}$ & ($0.029$) && 0.241$^{***}$ & ($0.033$) && 0.453$^{***}$ & ($0.031$)\\ 
  \hspace{0.75cm}Workload & 0.037 & ($0.035$) && 0.046 & ($0.032$) && 0.091$^{**}$ & ($0.034$)\\ 
  \hspace{0.75cm}Systemizing--Empathizing & -0.480$^{***}$ & ($0.046$) && -0.084 & ($0.045$) && -0.492$^{***}$ & ($0.049$)\\ 
  \hspace{0.75cm}Selectivity & -0.090$^{**}$ & ($0.028$) && -0.021 & ($0.028$) && -0.068$^{*}$ & ($0.028$) \\  
  \hspace{0.75cm}STEM & 0.239$^{***}$ & ($0.041$) && -0.257$^{***}$  & ($0.053$) && 0.114$^{*}$ & ($0.051$) \\
  \hspace{0.75cm}Homophily &&&& -0.397$^{***}$  & ($0.045$) \\ 
 \hspace{0.75cm}Quantitative GRE &&&&&&& 0.177$^{***}$ & ($0.047$) \vspace{0.4cm} \\ 
 
 Constant & -1.483$^{***}$ & ($0.016$) && -2.128$^{***}$ & ($0.021$) && -1.901$^{***}$ & ($0.020$) \\ 

\hline \hline 
\multicolumn{9}{l}{\textit{Note.} The coefficients are log odds ratios. The R syntax used to estimate Model A is as follows: Main\_Model\_I\enspace \textless \textendash \enspace \vspace{-2pt}}  \\ 

\multicolumn{9}{l} { glm(Left\_Field $\sim$ (Brilliance\_Orientation + Workload + Systemizing\_Empathizing + Selectivity + STEM) * Is\_Woman, \vspace{-2pt}} \\

\multicolumn{9}{l} {family = "binomial", data = Main\_Model\_I\_data). The cluster--robust standard errors and corresponding $p$ values were \vspace{-2pt}} \\

\multicolumn{9}{l} {generated as follows: coeftest(Main\_Model\_I, vcovCL(Main\_Model\_I, cluster = Main\_Model\_I\_data\$ORCID\_ID)). \vspace{-2pt}} \\

\multicolumn{9}{l} {$^{*}{\it p} < 0.05$; $^{**}{\it p} < 0.01$; $^{***}{\it p} < 0.001$}
\end{tabular} 
\end{table*} 

\clearpage


\begin{table*}
  \caption{Coefficients (and cluster--robust standard errors) from conditional logistic regressions that model the probability that an ORCID user {\bf joins a field} based on their gender (0 = man, 1 = woman) and the five field characteristics of primary interest (Model A), as well as homophily (Model B) or Quantitative GRE scores (Model C).} 
  \label{table:choice} 
\begin{tabular}{p{5.5cm}Sp{1.25cm}p{0.1cm}Sp{1.25cm}p{0.1cm}Sp{1cm}} 
 \hline \hline

& \multicolumn{2}{c}{\enspace \enspace Model A} && \multicolumn{2}{c}{\enspace \enspace Model B} && \multicolumn{2}{c}{\enspace \enspace  Model C} \\
& \multicolumn{2}{c}{\enspace \enspace Main Model II\vspace{1mm}} && 
\multicolumn{2}{c}{\enspace \enspace  Alternative: Homophily} && 
\multicolumn{2}{c}{\enspace \enspace Alternative: Quant GRE}\\
\hline
 
 Brilliance Orientation & 0.170$^{***}$ & ($0.014$) & & -0.596$^{***}$ & ($0.015$) & & -0.182$^{***}$ & ($0.016$) \\ 
 Workload & -0.800$^{***}$ & ($0.019$) & & -0.287$^{***}$ & ($0.024$) & & -0.248$^{***}$  & ($0.018$) \\ 
 Systemizing--Empathizing & -0.385$^{***}$ & ($0.017$) & & -0.438$^{***}$ & ($0.017$) & & -1.106$^{***}$  & ($0.020$) \\ 
 Selectivity & -0.905$^{***}$ & ($0.012$) & & -0.601$^{***}$  & ($0.012$) & & -1.043$^{***}$ & ($0.014$) \\ 
 STEM & 1.655$^{***}$ & ($0.027$)& & 0.045 & ($0.025$) & & 0.341$^{***}$  & ($0.025$) \\ 
 Homophily (log odds woman per field) & & & & -1.857$^{***}$  & ($0.015$) & & & \\
  Quantitative GRE & & & & & & & 1.726$^{***}$ & ($0.018$)\vspace{0.4cm}  \\
  
 Is Woman $\times$ \underline{\hspace{1.5cm}} \\
  \hspace{0.75cm}Brilliance Orientation & -1.011$^{***}$ & ($0.023$) & & -0.640$^{***}$  & ($0.025$) &  & -0.887$^{***}$ & ($0.027$) \\ 
  \hspace{0.75cm}Workload & 0.184$^{***}$ & ($0.035$) & & -0.117$^{**}$  & ($0.042$) & & 0.014 & ($0.035$) \\
  \hspace{0.75cm}Systemizing--Empathizing & 0.385$^{***}$ & ($0.029$) & & 0.293$^{***}$ & ($0.030$) & & 0.584$^{***}$ & ($0.034$)  \\ 
  \hspace{0.75cm}Selectivity & 0.267$^{***}$ & ($0.019$) & & 0.091$^{***}$ 
  & ($0.020$) & & 0.225$^{***}$ & ($0.023$) \\  
  \hspace{0.75cm}STEM & -1.221$^{***}$  & ($0.042$) & & -0.325$^{***}$ & ($0.042$) & & -0.929$^{***}$ & ($0.042$) \\
  \hspace{0.75cm}Homophily & & & & 0.781$^{***}$  & ($0.028$) & & & \\
  \hspace{0.75cm}Quantitative GRE & & & & & & & -0.338$^{***}$ & ($0.034$)\\
  
\hline \hline 
\multicolumn{9}{l}{\textit{Note.} The coefficients are log odds ratios. Conditional logit models do not estimate an intercept or coefficients for\vspace{-2pt}} \\ 

\multicolumn{9}{l}{variables that do not vary between the choices (in this case, academics' gender). The Stata syntax used to estimate Model \vspace{-2pt}} \\

\multicolumn{9}{l}{A is as follows: clogit\enspace Joined\_Field\enspace Is\_Woman\#\#(c.Brilliance\_Orientation\enspace c.Workload\enspace c.Systemizing\_Empathizing\enspace \vspace{-2pt}} \\

\multicolumn{9}{l}{ c.Selectivity\enspace STEM),\enspace group(\_caseid)\enspace vce(cluster ORCID\_ID). The variable \_caseid marks each ``choice'': a group of 29  \vspace{-2pt}} \\

\multicolumn{9}{l}{observations consisting of 28 unchosen fields and 1 chosen field. $^{*}{\it p} < 0.05$; $^{**}{\it p} < 0.01$; $^{***}{\it p} < 0.001$}   
\end{tabular} 
\end{table*} 

\clearpage


\begin{table*}
  \caption{Coefficients (and cluster--robust standard errors) from logistic regressions that predict the gender of an ORICD user {\bf transitioning between fields} (0 = man, 1 = woman) based on the five field characteristics of primary interest (Model A), as well as homophily (Model B) or Quantitative GRE scores (Model C).} 
  \label{table:flow} 
\begin{tabular}{p{4cm}Sp{1.25cm}p{0.1cm}Sp{1.25cm}p{0.1cm}Sp{1cm}} 
 \hline \hline

& \multicolumn{2}{c}{\enspace \enspace Model A} && \multicolumn{2}{c}{\enspace \enspace Model B} && \multicolumn{2}{c}{\enspace  Model C} \\
& \multicolumn{2}{c}{\enspace \enspace Main Model III\vspace{1mm}} && 
\multicolumn{2}{c}{\enspace \enspace  Alternative: Homophily} && 
\multicolumn{2}{c}{\enspace Alternative: Quant GRE}\\
\hline

 Brilliance Orientation & -0.507$^{***}$ & ($0.040$) && -0.104$^{*}$ & ($0.045$) && -0.366$^{***}$ & ($0.045$)\\ 
 Workload & 0.123$^{***}$ & ($0.034$) && 0.049 & ($0.032$) && 0.087$^{*}$  & ($0.034$) \\ 
 Systemizing--Empathizing & 0.009 & ($0.048$) && 0.015 & ($0.046$) && 0.147$^{**}$  & ($0.055$) \\ 
 Selectivity & 0.074$^{**}$ & ($0.028$) && -0.051 & ($0.030$) && 0.027  & ($0.030$) \\ 
 STEM & -0.440$^{***}$ & ($0.035$) && -0.048  & ($0.043$) && -0.319$^{***}$ & ($0.040$) \\ 
 Homophily &&&& 0.678$^{***}$ & ($0.040$) && \\ 
 Quantitative GRE &&&&&&& -0.348$^{***}$  & ($0.042$)  \\ 
 Constant & 0.103 & ($0.064$) && 0.131$^{*}$ & ($0.064$) && 0.038  & ($0.067$) \\ 
  
\hline \hline 
\multicolumn{9}{l}{\textit{Note.} The coefficients are log odds ratios. The coefficients for the 29 field indicator variables are omitted. The R \vspace{-2pt}}\\ 

\multicolumn{9}{l}{syntax used to estimate Model A is as follows: Main\_Model\_III\enspace \textless \textendash \enspace glm(Is\_Woman $\sim$ Brilliance\_Orientation + \vspace{-2pt}}  \\ 

\multicolumn{9}{l} { Workload + Systemizing\_Empathizing + Selectivity + STEM + Source\_Field, family = "binomial", data = \vspace{-2pt}} \\

\multicolumn{9}{l} {Main\_Model\_III\_data). Source\_Field is a 30-level factor variable that is implemented as 29 indicator variables.  \vspace{-2pt}} \\

\multicolumn{9}{l} {The cluster--robust SEs and corresponding $p$ values were generated as follows: coeftest(Main\_Model\_III,  \vspace{-2pt}} \\

\multicolumn{9}{l}{vcovCL(Main\_Model\_III, cluster = Main\_Model\_III\_data\$ORCID\_ID)). $^{*}{\it p} < 0.05$; $^{**}{\it p} < 0.01$; $^{***}{\it p} < 0.001$}  
\end{tabular} 
\end{table*} 

\clearpage

\section{\large{Supplementary Text: \\ORCID Data Processing Details}}\label{section:ORCID}
\vspace{0.5cm}

\twocolumngrid

We provide a narrative summary of our data processing steps to accompany our publicly available scripts at {\href{https://github.com/kennyjoseph/ORCID_career_flows}{https://github.\\com/kennyjoseph/ORCID\_career\_flows}}.

ORCID users have the option to selectively make their information public. This opt-in public information is released annually in an aggregated ORCID public data file~\cite{ORCID} and also made available on demand to ORCID member institutions. The most recent version of ORCID public data were accessed on September 18, 2019, using the member institution status of the University of Colorado Boulder.

Our goal is to identify the subset of researchers whom we can confidently place in particular fields of study, over time or career stage, and whose first and last names provide us with a high-confidence association with a gender. However, ORCID does not provide information about its users' fields of study or gender. As a consequence, not all ORCID profiles could be analyzed, so this section describes our data cleaning and inclusion processes in detail. We begin with a dataset of 6,485,785 public profiles of researchers on ORCID. Only 1,983,632 researchers have one or more listed affiliations, and among them we observe a total of 5,307,437 affiliations.

{\bf Step 1.} In the first step, we perform basic filtering to remove incomplete or out-of-scope affiliations. Specifically, we filter out three types of affiliations. First, we remove any affiliations where the researcher’s name was not provided, as we use names to estimate an associated gender. Second, we filter out affiliations where the department name was not provided, as we use this information to identify the academic field associated with the affiliation. Finally, we remove any affiliations where neither a career stage (e.g., ``postdoc'') nor a date was provided, as we use this information to determine the ordering of affiliations. (The order of affiliations will later be used to identify field switches and determine the source and destination fields involved in a switch.) After these three filters have been applied, we retain 3,988,331 affiliations among 1,287,228 researchers.

{\bf Step 2.} In the second step, we fill in variables of interest using three algorithms: one to determine the role/job position affiliated with each affiliation (e.g., PhD candidate, professor; see Section~\ref{sub:stages}), one to identify the academic field associated with each affiliation (see Section~\ref{sub:fields}), and one to determine the gender that is culturally associated with researchers' names (see Section~\ref{sub:genders}). Details on the algorithms themselves can be found in the sections that follow, but here we summarize their outputs. We remove affiliations that (i) match a field that is not among the 30 surveyed fields, or (ii) match multiple fields, retaining 1,274,087 affiliations (685,649 researchers). Among these, we are able to find a high-confidence inferred gender via a cultural consensus algorithm for 1,027,250 affiliations (550,961 researchers; see Supplementary Table~\ref{table:breakdown_by_field} for a breakdown by field). We note that these were the data that were used to calculate the homophily variables for the relevant analyses in the main text.

\begin{table*}
\small
\centering
\captionsetup{justification=centering}
\caption{Breakdown of ORCID users in our dataset by field and gender (proportions in parentheses).}
\label{table:breakdown_by_field}
\begin{tabular}{l>{\raggedleft}p{1.2cm}>{\raggedleft}p{1.0cm}>{\raggedleft}p{.8cm}>{\raggedleft}p{1.0cm}>{\raggedleft}p{.8cm}p{.5cm}>{\raggedleft}p{1.2cm}>{\raggedleft}p{1.0cm}>{\raggedleft}p{.8cm}>{\raggedleft}p{1.0cm}>{\raggedleft\arraybackslash}p{.8cm}} 
  \hline \hline
 &  \multicolumn{5}{c}{{\bf Everyone, regardless of whether they}} && \multicolumn{5}{c}{{\bf Those who switched out of}} \\
  &  \multicolumn{5}{c}{{\bf ever switched fields}} && \multicolumn{5}{c}{{\bf these fields at some point}} \\
\hline
   Field & \raggedleft Total & \multicolumn{2}{c}{ \hspace{9pt} Men }   & \multicolumn{2}{c}{ \hspace{7pt} Women } && \raggedleft Total  & \multicolumn{2}{c}{ \hspace{9pt} Men } & \multicolumn{2}{c}{ \hspace{7pt} Women }  \\
  \hline
Anthropology & 7283 & 3567 & (0.49) & 3716 & (0.51) && 1230 & 615 & (0.50) & 615 & (0.50) \\
Archaeology & 3019 & 1722 & (0.57) & 1297 & (0.43) && 524 & 278 & (0.53) & 246 & (0.47) \\
Art History & 2194 & 863 & (0.39) & 1331 & (0.61) && 358 & 141 & (0.39) & 217 & (0.61) \\
Astronomy &  51 &  38 & (0.75) &  13 & (0.25) &&  12 &   8 & (0.67) &   4 & (0.33) \\
Biochemistry & 16213 & 10280 & (0.63) & 5933 & (0.37) && 2184 & 1481 & (0.68) & 703 & (0.32) \\
Chemistry & 61290 & 42780 & (0.70) & 18510 & (0.30) && 7430 & 5432 & (0.73) & 1998 & (0.27) \\
Classics & 1685 & 897 & (0.53) & 788 & (0.47) && 382 & 197 & (0.52) & 185 & (0.48) \\
Communications & 13608 & 6838 & (0.50) & 6770 & (0.50) && 1994 & 1068 & (0.54) & 926 & (0.46) \\
Comparative Literature & 686 & 317 & (0.46) & 369 & (0.54) && 174 &  80 & (0.46) &  94 & (0.54) \\
Computer Science & 33952 & 27294 & (0.80) & 6658 & (0.20) && 4332 & 3541 & (0.82) & 791 & (0.18) \\
Earth Sciences & 10385 & 7157 & (0.69) & 3228 & (0.31) && 1074 & 741 & (0.69) & 333 & (0.31) \\
Economics & 35821 & 23341 & (0.65) & 12480 & (0.35) && 2806 & 1833 & (0.65) & 973 & (0.35) \\
Education & 48301 & 22400 & (0.46) & 25901 & (0.54) && 5605 & 3116 & (0.56) & 2489 & (0.44) \\
Engineering & 154381 & 126400 & (0.82) & 27981 & (0.18) && 12422 & 10167 & (0.82) & 2255 & (0.18) \\
English Literature & 12572 & 5826 & (0.46) & 6746 & (0.54) && 2524 & 1184 & (0.47) & 1340 & (0.53) \\
Evolutionary Biology & 2259 & 1313 & (0.58) & 946 & (0.42) && 120 &  63 & (0.53) &  57 & (0.47) \\
History & 17251 & 10254 & (0.59) & 6997 & (0.41) && 2752 & 1627 & (0.59) & 1125 & (0.41) \\
Linguistics & 6578 & 3029 & (0.46) & 3549 & (0.54) && 1315 & 655 & (0.50) & 660 & (0.50) \\
Mathematics & 29864 & 22605 & (0.76) & 7259 & (0.24) && 6331 & 4809 & (0.76) & 1522 & (0.24) \\
Middle Eastern Studies & 370 & 243 & (0.66) & 127 & (0.34) &&  87 &  51 & (0.59) &  36 & (0.41) \\
Molecular Biology & 6687 & 4066 & (0.61) & 2621 & (0.39) && 754 & 507 & (0.67) & 247 & (0.33) \\
Music Theory \& Composition & 878 & 521 & (0.59) & 357 & (0.41) && 125 &  72 & (0.58) &  53 & (0.42) \\
Neuroscience & 8721 & 5092 & (0.58) & 3629 & (0.42) && 791 & 460 & (0.58) & 331 & (0.42) \\
Philosophy & 12185 & 8151 & (0.67) & 4034 & (0.33) && 2459 & 1583 & (0.64) & 876 & (0.36) \\
Physics & 59215 & 48410 & (0.82) & 10805 & (0.18) && 9342 & 7791 & (0.83) & 1551 & (0.17) \\
Political Science & 12331 & 7903 & (0.64) & 4428 & (0.36) && 1659 & 1033 & (0.62) & 626 & (0.38) \\
Psychology & 39262 & 17032 & (0.43) & 22230 & (0.57) && 4564 & 2178 & (0.48) & 2386 & (0.52) \\
Sociology & 11812 & 6206 & (0.53) & 5606 & (0.47) && 1799 & 963 & (0.54) & 836 & (0.46) \\
Spanish Literature & 1567 & 712 & (0.45) & 855 & (0.55) && 350 & 146 & (0.42) & 204 & (0.58) \\
Statistics & 7706 & 5229 & (0.68) & 2477 & (0.32) && 1186 & 844 & (0.71) & 342 & (0.29) \\

 \hline \hline
\vspace{-7pt} \\
\multicolumn{12}{l}{{\it Note.} The two sets of statistics above correspond to Step 2 and 3 in ORCID data processing sequence described in Section}\\
\multicolumn{12}{l}{\ref{section:ORCID}. However, the numbers in the ``Total'' columns will add up to more than the numbers provided in Section~\ref{section:ORCID} because users}\\
\multicolumn{12}{l}{can be affiliated with more than one field.}

\end{tabular}
\end{table*}

{\bf Step 3.} In the third step, we take the 1,027,250 affiliations among 550,961 researchers and identify pairs of affiliations that indicate that a researcher switched from one of the 30 surveyed fields to another (see Section~\ref{sub:transitions}). To do so, we order each researcher's affiliations using the roles identified in Step 2 and/or the start date of the affiliation (for an illustration, see Supplementary Fig.~\ref{Fig:ORCIDannotations}). Whenever the fields associated with consecutive affiliations are different, we record a transition as having occurred from one affiliation's field to the next affiliation's field. If an individual changes fields multiple times, each transition is recorded as a separate transition, but the transitive transition (e.g., the first field to the third field) is not recorded. In total, we are able to identify 78,798 transitions among 61,108 researchers (see Supplementary Table~\ref{table:breakdown_by_field}), averaging 1.3 transitions per person among the 11.1\% of researchers with observable transitions.

\subsection{Determining Career Stages}\label{sub:stages}

Each ORCID affiliation has an associated {\bf role} field, which we use to identify the career stage associated with that affiliation. Due to the fact that the ORCID userbase spans many languages and academic traditions, we used a set of regular expressions to coarse-grain each affiliation into one of the following academic career stages: bachelor’s degree, master’s degree, PhD, postdoctoral researcher, and professor/department head. In the event that the text in an affiliation matches multiple stages, we select the highest ranking role. In the event that there is no match, we give that affiliation a blank role, since affiliations that have no role but nevertheless have a date that can be placed in sequence with other affiliations are still useful in our analysis. 

The regular expressions were accumulated recursively: After every iteration of matching, we manually identified the most commonly missed expressions among unmatched roles, and then added a corresponding regular expression. We stopped once the inclusion of additional regular expressions did not substantially improve our data coverage. The complete set of regular expressions has been made publicly available. A breakdown of the career stages identified in our dataset via this algorithm can be found in Supplementary Table~\ref{table:careers}.

\begin{table}
 
      \caption{Breakdown of career stages in the dataset.}
    \label{table:careers}
    \begin{tabular}{l>{\raggedleft}r>{\raggedleft\arraybackslash}r}
    \hline\hline
    \textbf{Role}             & \textbf{Number of} & \textbf{\hspace{5pt}\% of} \\ 
    & \textbf{Affiliations} & \textbf{\hspace{5pt}Data}\\
    \hline
    Bachelor’s Degree         & 157330                      & \hspace{5pt}12.1\%             \\ 
    Master’s Degree           & 183313                      & \hspace{5pt}14.1\%             \\ 
    PhD                       & 306434                      & \hspace{5pt}23.6\%             \\ 
    Postdoctoral Researcher   & 47866                       & \hspace{5pt}3.7\%              \\ 
    Professor/Department Head & 253404                      & \hspace{5pt}19.5\%             \\ 
    Other/None                & 348705                      & \hspace{5pt}26.9\%             \\ \hline\hline

    \end{tabular}

\end{table}

\subsection{Determining Academic Fields}\label{sub:fields}

\begin{table}[t!]
	\footnotesize
    \caption{Fields and the corresponding terms used for matching ORCID affiliation strings.}
    \label{table:fields}
    \begin{tabular}{|p{2.8cm}|p{5.3cm}|}
    \hline
    \textbf{Fields}        & \textbf{Accepted expressions}\\ \hline\hline
    Anthropology           & anthropology\\ \hline
    Archaeology            & archaeology \\ \hline
    Art History            & art history; history of art                                                 \\ \hline
    Astronomy              & astronomy   \\ \hline
    Biochemistry           & biochemistry\\ \hline
    Chemistry              & chemistry   \\ \hline
    Classics               & classics; classical literature; classical humanities                        \\ \hline
    Communications         & communications; communication sciences; communication studies; communication \\ \hline
    Comparative Literature & comparative literature                                                     \\ \hline
    Computer Science       & computer science; algorithms; computing; informatics                          \\ \hline
    Earth Sciences         & earth sciences; earth science; physical geography; oceanography; atmospheric sciences; volcano                                                                                                                                                                                                                                                                                                      \\ \hline
    Economics              & economics; economic; econometrics; finance; economy                           \\ \hline
    Education              & education; pedagogy                                                         \\ \hline
    Engineering            & engineering; ingegneria; e.e.; e.c.e.; ingenierã a; cybernetics; telecommunication; telecommunications; telecommunication studies; electrical engineering; chemical engineering; electrical and computer engineering; biochemical engineering; biological engineering; neuroengineering; musical engineering; statistical engineering; physical engineering \\ \hline
    English Literature     & english literature; english                                                 \\ \hline
    Evolutionary Biology   & evolutionary biology                                                       \\ \hline
    History                & history     \\ \hline
    Linguistics            & linguistics; linguistic                                                    \\ \hline
    Mathematics            & mathematics; math; geometry; algebra; number                                   \\ \hline
    Middle Eastern Studies & middle eastern studies; middle east                                        \\ \hline
    Molecular Biology      & molecular biology                                                          \\ \hline
    Music Theory \& Comp.                  & music theory; musical composition; musicology; composition                   \\ \hline
    Neuroscience           & neuroscience\\ \hline
    Philosophy             & philosophy  \\ \hline
    Physics                & physics     \\ \hline
    Political Science      & political science; political sciences; politics; science politique; politology\\ \hline
    Psychology             & psychology; psychological; psicologia; psicología                           \\ \hline
    Sociology              & sociology; sociological; sociologie                                          \\ \hline
    Spanish Literature     & spanish literature; spanish                                             \\ \hline
    Statistics             & statistics; statistical sciences                                           \\ \hline
    \end{tabular}
\end{table}

Each ORCID affiliation has an associated {\bf department name} field (hereafter, department). Only those affiliations that can be confidently linked to one and only one of the 30 surveyed academic fields~\cite{leslie2015expectations} can be used in our analysis, so we now describe the procedure used to match user-provided departments with surveyed academic fields. There are three steps in our approach: translation, matching, and multi-field affiliation removal.

{\bf Step 1: Translation.} In the translation step, we use a list of common academia-related English words to determine whether or not an affiliation's department is in English. We then translated the 160,156 non-English unique department names into English using Google Translate.

{\bf Step 2: Matching.} In the matching step, we first construct, for each of the 30 surveyed fields, a list of expressions and subfields that are associated with that field. For example, {\it sociology} or {\it sociological} could both map to the field of sociology. Supplementary Table~\ref{table:fields} provides a complete list of fields and their corresponding expressions.  Note that we retained certain popular non-English terms, as there were some instances in which certain terms were not translated (they were considered to be misspellings by the translation algorithm).

We also assembled a so-called {\it denylist} of scientific fields that are prominent in the ORCID data but that are not in our survey data. Constructing this list helped identify affiliations in fields that were clearly defined but outside the scope of our study; researchers who will use this dataset in the future might find these fields useful. 

To construct the terms in Supplementary Table~\ref{table:fields} and our denylist, we used a recursive approach, using existing resources from Wikipedia and the U.S. National Science Foundation to determine initial lists of terms, and then repeatedly inspecting department names that appeared multiple times in our dataset to ensure coverage of our lists.  

To validate the output of the matching step, we hand-checked the fields assigned to departments from a stratified random sample, consisting of  25\% matched and in-sample affiliations, 25\% unmatched affiliations, and 50\% matched but out of sample (denylist) affiliations. Each affiliation was assigned to two of the four authors of the study for annotation. We used disagreements between annotators and the matching step to improve the expressions and denylist. In total, 660 affiliations were checked by hand by at least two authors, and all disagreements were discussed by all authors.

With the final set of terms associated with each field (see Supplementary Table~\ref{table:fields}) and the denylist, we use a simple rule-based algorithm to match affiliations to fields. The algorithm works as follows. 

First, it splits each affiliation string on common separators (e.g., commas, "and") into candidate match objects. For instance, consider the fake and implausible department (with intentional misspelling) {\it Advanced Chemical Engineering and Histroy/History of Art}. Based on the matching terms, this string contains three candidate match objects: (i) ``Advanced Chemical Engineering'', (ii) ``Histroy'', and (iii) ``History of Art''.  

Then, for each candidate match object, we check whether it is an exact match to any term associated with a surveyed field or a field on the denylist. If so, we have identified the field associated with the candidate match object, and move to the next candidate match object. For example, ``History of Art'' matches a term in Supplementary Table~\ref{table:fields} associated with the field Art History.  

If a candidate match object matches no known terms, we check for fuzzy string matches with an edit distance of 3 or less. If there are any such fuzzy matches, we select the one with the smallest edit distance, and break ties by choosing the longest of the matched strings. For example, the candidate match object ``Histroy'' has an edit distance of 1 to the match term ``History,'' which is linked to the academic field History.  

Finally, we check for exact matches and/or fuzzy matches in subsets of the candidate match strings.  For example, there is an exact match to ``Chemical Engineering'', a term in Supplementary Table~\ref{table:fields}, within the candidate match object “Advanced Chemical Engineering”. In such a case, we identify this field as a match, remove the relevant substring, and then continue recursively (i.e., try to match the remaining substring “Advanced”). 

This field matching method identifies zero, one, or more academic fields associated with each affiliation. In total, 59.1\% of the in-sample affiliations matched \emph{exactly} to one or more of the terms in Supplementary Table~\ref{table:fields}, with another 33.7\% within an edit distance of 2 to one or more of those terms. Thus, 92.8\% of the affiliations we identified as representing one of the 30 fields in our survey data were linked to that field either because they were an exact match to a term in Supplementary Table~\ref{table:fields} or were a slight misspelling of one of those terms.

{\bf Step 3: Multi-field affiliation removal.} The matching procedure has the potential to associate multiple fields with a single affiliation. In such cases, the affiliation was conservatively removed from further consideration in order to avoid ambiguities.

\subsection{Determining Region} \label{sub:region}

Each ORCID affiliation has an associated {\it ISO 3166-1 alpha-2} country code, allowing us to test the extent to which field transitions in different parts of the world show the same patterns as those in the global dataset. Supplementary Table~\ref{table:region} lists which countries are assigned to which regions, and the number of transitions to organizations in countries in that region.

\begin{table*}[t]
    \caption{Regions represented in the ORCID dataset, alongside the corresponding countries.}
    \label{table:region}
    \footnotesize
    \begin{tabular}{|p{2.1cm}|l|p{13.3cm}|}
    \hline
    \textbf{Region} & \textbf{N. of Transitions}       & \textbf{Countries}\\ \hline\hline
     Europe & 30718 & Albania, Andorra, Austria, Belarus, Belgium, Bosnia and Herzegovina, Bulgaria, Channel Islands, Croatia, Czech Republic, Denmark, Estonia, Faroe Islands, Finland, France, Germany, Gibraltar, Greece, Greenland, Guernsey, Holy See (Vatican City State), Hungary, Iceland, Ireland, Isle of Man, Italy, Jersey, Kosovo, Latvia, Liechtenstein, Lithuania, Luxembourg, Malta, Monaco, Montenegro, Netherlands, Norway, Poland, Portugal, Republic of Moldova, Romania, Russian Federation, San Marino, Serbia, Slovakia, Slovenia, Spain, Svalbard and Jan Mayen, Sweden, Switzerland, The Former Yugoslav Republic of Macedonia, Ukraine, United Kingdom of Great Britain and Northern Ireland\\ \hline   
    Northern America  & 20854 & Canada, United States of America \\ \hline
    Asia & 11540 & Afghanistan, Armenia, Azerbaijan, Bahrain, Bangladesh, Bhutan, Brunei Darussalam, Cambodia, China, China, Hong Kong Special Administrative Region, China, Macao Special Administrative Region, Cyprus, Democratic People's Republic of Korea, Georgia, India, Indonesia, Iran (Islamic Republic of), Iraq, Israel, Japan, Jordan, Kazakhstan, Kuwait, Kyrgyzstan, Lao People's Democratic Republic, Lebanon, Malaysia, Maldives, Mongolia, Myanmar, Nepal, Occupied Palestinian Territory, Oman, Pakistan, Philippines, Qatar, Republic of Korea, Saudi Arabia, Singapore, Sri Lanka, Syrian Arab Republic, Taiwan, Tajikistan, Thailand, Timor-Leste, Turkey, Turkmenistan, United Arab Emirates, Uzbekistan, Viet Nam, Yemen \\ \hline
    Latin America and the Caribbean  & 11215 & Anguilla, Antigua and Barbuda, Argentina, Aruba, Bahamas, Barbados, Belize, Bolivia (Plurinational State of), Brazil, Cayman Islands, Chile, Colombia, Costa Rica, Cuba, Curacao, Dominica, Dominican Republic, Ecuador, El Salvador, French Guiana, Grenada, Guadeloupe, Guatemala, Guyana, Haiti, Honduras, Jamaica, Martinique, Mexico, Montserrat, Netherlands Antilles, Nicaragua, Panama, Paraguay, Peru, Puerto Rico, Saint Kitts and Nevis, Saint Lucia, Saint Vincent and the Grenadines, Suriname, Trinidad and Tobago, United States Virgin Islands, Uruguay, Venezuela (Bolivarian Republic of), Virgin Islands (British) \\ \hline
    Oceania  & 2443 & American Samoa, Australia, Fiji, French Polynesia, Guam, Kiribati, Marshall Islands, Micronesia (Federated States of), Nauru, New Caledonia, New Zealand, Papua New Guinea, Pitcairn Islands, Samoa, Solomon Islands, Tonga, Tuvalu, Vanuatu, Wallis and Futuna \\ \hline
    Africa  & 2026 & Algeria, Angola, Benin, Botswana, Burkina Faso, Burundi, Cameroon, Cape Verde, Central African Republic, Chad, Comoros, Congo, Côte d'Ivoire, Democratic Republic of the Congo, Djibouti, Egypt, Equatorial Guinea, Eritrea, Ethiopia, Gabon, Gambia, Ghana, Guinea, Guinea-Bissau, Kenya, Lesotho, Liberia, Libyan Arab Jamahiriya, Madagascar, Malawi, Mali, Mauritania, Mauritius, Mayotte, Morocco, Mozambique, Namibia, Niger, Nigeria, Réunion, Rwanda, Sao Tome and Principe, Senegal, Seychelles, Sierra Leone, Somalia, South Africa, South Sudan, Sudan, Swaziland, Togo, Tunisia, Uganda, United Republic of Tanzania, Western Sahara, Zambia, Zimbabwe\\ \hline

\multicolumn{3}{l}{\vspace{-4pt}} \\
\multicolumn{3}{l}{\textit{Note.} Regions are listed in decreasing order of transition counts. Transitions to institutions in Oceania and Africa were not analyzed separately due to}   \\
\multicolumn{3}{l}{insufficient statistical power, but are included for completeness.}  \\

    \end{tabular}
\end{table*}

\subsection{Associating Names with Gender}\label{sub:genders}

\begin{table*}[t]
    \captionsetup{justification=centering}
    \caption{Links to the 44 lists and databases used to \\ infer cultural name--gender associations.}
    \label{table:gender}
    \footnotesize
    \begin{tabular}{|l|l|}
    \hline 
    \textbf{Linked Source}             & \textbf{Type} \\ \hline \hline
\href{http://www.statistik.at/web_de/statistiken/menschen_und_gesellschaft/bevoelkerung/geborene/vornamen/index.html}{Austria (AUT)} & Counts \\ \hline
\href{https://statbel.fgov.be/nl/themas/bevolking/namen-en-voornamen/voornamen-van-meisjes-en-jongens#panel-12}{Belgium (BEL)} & Counts \\ \hline
\href{https://www.bfs.admin.ch/bfs/de/home/statistiken/bevoelkerung/geburten-todesfaelle/vornamen-schweiz.html}{Switzerland (CHE)} & Counts \\ \hline
\href{https://github.com/OpenGenderTracking/globalnamedata/tree/master/assets}{UK (GBR)} & Counts \\ \hline
\href{https://www.avoindata.fi/data/en_GB/dataset/none}{Finland (FIN)} & Counts \\ \hline
\href{https://pxweb.stat.si/SiStatDb/pxweb/en/10_Dem_soc/10_Dem_soc__05_prebivalstvo__46_Imena_priimki__06_05X10_imena_priimki/?tablelist=true}{Slovenia (SVN)} & Counts \\ \hline
\href{https://www.ine.es/dyngs/INEbase/es/operacion.htm?c=Estadistica_C&cid=1254736177009&menu=resultados&secc=1254736195454&idp=1254734710990}{Spain (ESP)} & Counts \\ \hline
\href{https://www.beliebte-vornamen.de/jahrgang}{Germany (DEU)} & Dictionary \\ \hline
\href{https://www.ssa.gov/oact/babynames/limits.html}{United States (USA)} & Counts \\ \hline
\href{https://www.insee.fr/fr/statistiques/2540004#documentation}{France (FRA)} & Counts \\ \hline
\href{https://data.sa.gov.au/data/dataset/popular-baby-names/resource/534d13f2-237c-4448-a6a3-93c07b1bb614}{Australia (AUS)} & Counts \\ \hline
\href{https://data.ontario.ca/dataset?q=baby+names}{Canada (CAN)} & Counts \\ \hline
\href{https://statbank.cso.ie/px/pxeirestat/Database/eirestat/Irish%20Babies%20Names/Irish%20Babies%20Names_statbank.asp?SP=Irish%20Babies%20Names&Planguage=0}{Ireland (IRL)} & Counts \\ \hline
\href{https://www.ssb.no/en/statbank/table/10467/}{Norway (NOR)} & Counts \\ \hline
\href{https://www.dia.govt.nz/diawebsite.nsf/wpg_URL/Services-Births-Deaths-and-Marriages-Most-Popular-Male-and-Female-First-Names?OpenDocument}{New Zealand (NZL)} & Top-100-Counts \\ \hline
\href{https://www.nyilvantarto.hu/archiv_honlap/kozos/index.php?k=statisztikai_adatok_lakossagi_legujsznevek_hu_archiv}{Hungary (HUN)} & Top-100-Counts \\ \hline
\href{https://www.scb.se/hitta-statistik/statistik-efter-amne/befolkning/amnesovergripande-statistik/namnstatistik/}{Sweden (SWE)} & Top-100-Counts \\ \hline
\href{http://www.turkstat.gov.tr/PreTablo.do?alt_id=1059}{Turkey (TUR)} & Top-100-Ranks \\ \hline
\href{https://www.behindthename.com/top/lists/poland}{Poland (POL)} & Top-100-Counts \\ \hline
\href{https://www.behindthename.com/top/lists/russia-moscow}{Russia (RUS)} & Top-50-Counts \\ \hline
\href{https://www.behindthename.com/top/lists/czech}{Czech Republic (CZE)} & Top-100-Counts \\ \hline
\href{https://www.behindthename.com/top/lists/portugal}{Portugal (PRT)} & Top-100-Counts \\ \hline
\href{https://www.behindthename.com/top/lists/bosnia}{Bosnia and Herzegovina (BIH)} & Top-100-Counts \\ \hline
\href{https://www.behindthename.com/top/lists/romania}{Romania (ROU)} & Top-50-Counts \\ \hline
\href{https://www.behindthename.com/top/lists/lithuania}{Lithuania (LTU)} & Top-20-Ranks \\ \hline
\href{https://www.behindthename.com/top/lists/denmark}{Denmark (DNK)} & Top-50-Counts \\ \hline
\href{https://www.behindthename.com/top/lists/israel}{Israel (ISR)} & Top-100-Counts \\ \hline
\href{https://www.behindthename.com/top/lists/chile}{Chile (CHL)} & Top-100-Counts \\ \hline
\href{https://www.behindthename.com/top/lists/netherlands}{The Netherlands (NLD)} & Top-500-Percents \\ \hline
\href{https://www.behindthename.com/top/lists/italy}{Italy (ITA)} & Top-200-Percents \\ \hline
\href{https://www.behindthename.com/top/lists/iceland}{Iceland (ISL)} & Top-50-Percents \\ \hline
\href{https://www.behindthename.com/top/lists/croatia}{Croatia (HRV)} & Top-100-Ranks \\ \hline
\href{https://www.behindthename.com/top/lists/mexico}{Mexico (MEX)} & Top-100-Ranks \\ \hline
\href{https://sites.google.com/site/facebooknamelist/namelist}{Facebook-Name-List} & Counts \\ \hline
\href{https://www.cs.cmu.edu/Groups/AI/util/areas/nlp/corpora/names/}{Kantrowitz-Data} & Dictionary \\ \hline
\href{https://www.ine.es/dyngs/INEbase/es/operacion.htm?c=Estadistica_C&cid=1254736177009&menu=resultados&secc=1254736195454&idp=1254734710990}{INE} & Dictionary \\ \hline
\href{https://osf.io/bt9ya/files/}{Holman-Genderize-Arxiv} & Score \\ \hline
\href{http://abel.lis.illinois.edu/cgi-bin/ethnea/search.py?Fname=Andrea+Maria&Lname=Ventura}{Genni-Ethnea} & Dictionary \\ \hline
\href{https://www.behindthename.com/names}{Behind-the-Names} & Dictionary \\ \hline
\href{https://www.wikidata.org/wiki/Wikidata:WikiProject_Names/lists/given_names/1}{Wikidata-Names} & Dictionary \\ \hline
\href{http://whgi.wmflabs.org/}{WHGI} & Counts \\ \hline
\href{https://www.kaggle.com/heesoo37/120-years-of-olympic-history-athletes-and-results}{Olympic-History} & Counts \\ \hline
\href{ftp://ftp.musicbrainz.org/pub/musicbrainz/data/json-dumps}{MusicBrainz} & Counts \\ \hline
\href{https://github.com/lead-ratings/gender-guesser/tree/master/gender_guesser/data}{Jorg-Michael-Data} & Dictionary \\ \hline
    \end{tabular}
\end{table*}

ORCID neither collects nor infers gender information. Thus, we inferred the extent to which each user's first and last names are culturally associated with different gender labels. This inferential process was guided by the theoretical framework of cultural consensus models~\cite{batchelder1988test}, which do not purport to identify the ``true'' gender label of an individual but instead measure the consensus across multiple viewpoints. In other words, this method does not ask ``What is Jane Doe's gender?'' but rather ``What is the likelihood that someone with the name `Jane Doe' is thought to be a woman?'' In this way, the inference algorithm attempts to estimate how an individual is likely to be perceived based on their name.

Our consensus-based gender inference algorithm computes the Bayesian posterior probability that a person’s name is culturally understood to be the name of a woman (or complementarily, a man) based on data from 44 different sources, ranging from the U.S. Social Security Administration’s names database to a list of the world’s Olympic athletes (see Supplementary Table \ref{table:gender}). 

Names that did not appear in any of the 44 reference datasets were submitted to Genni~\cite{torvik2016ethnea}, a service that takes into account the perceived ethnicity of first and last names to improve estimates of the cultural associations between first names and gender. 

Finally, names with posterior probabilities or Genni scores of $\geq 0.9$ and $\leq 0.1$ were associated with the labels {\it woman} and {\it man}, respectively. Conservatively, the 20.5\% of names with scores between 0.1 and 0.9 were not included in our analyses.

\subsection{Identifying Field Transitions}\label{sub:transitions}

The most critical element of the algorithm that identifies field transitions is the one that {\it orders} the affiliations of a given researcher.  As noted above, we consider only those affiliations with either a clear academic role or a start date. When all of a researcher's affiliations have a date, ordering is trivial. In fact, because 96.1\% of the affiliations had a start date, they were easily ordered in the vast majority of cases. In the remaining cases, when all of a researcher’s affiliations are associated with one of the clear academic career stages considered in this paper (see Section~\ref{sub:stages}), and a researcher has no more than one affiliation per stage, we assumed an order of bachelor’s degree $\to$ master’s degree $\to$ PhD $\to$ postdoctoral researcher $\to$ professor/department head. In this case, again, ordering is trivial.

The only difficult remaining cases are those researchers whose affiliations are a mixture of dates without career stages and career stages without dates. In this case, we used a simple algorithm that attempts to interleave affiliations. The algorithm takes advantage of any cases where the researcher lists both a date and a career stage, using such affiliations as an anchor to sort the other affiliations by dates and career stages, again under the same ordered career stage assumption as above. Implementations of these algorithms have been made publicly available.

\clearpage
\onecolumngrid
\section{\large{Supplementary Text: \\Robustness to Sampling Bias in the ORCID Data}}
\label{section:simulations}
\vspace{0.5cm}

\twocolumngrid

ORCID usage varies by field, and not all users opt to make their information public. This raises the question of whether the public ORCID dataset can indeed be used to produced unbiased estimates of the relation between field attributes and women's and men's transitions between fields.

In this section, we describe a set of numerical experiments with synthetic data. In each of these experiments, we first bias and censor the data in ways that ORCID data might also be biased and censored. We then test whether these manipulations detract from our ability to produce unbiased estimates of the ``true'' relationships. In other words, our goal is to map out the circumstances under which our analyses are robust to potential problems with ORCID data. All synthetic data and the code used to analyze them are available in the GitHub repository for this paper. 

Our bias scenarios correspond to answering the following questions. 
\begin{enumerate}
    \item What if we had complete data for all transitions between fields?
    \item What if only 10\% of transitions were in our dataset, across fields?
    \item What if only 10\% of transitions were in our dataset, but due to differences in ORCID usage by field, the percentage of transitions per field similarly varied between 0\% and 20\%? 
    \item What if, in addition to observing only 0\% to 20\% of transitions for each field, there were also variable bias in how popular ORCID adoption is {\it by gender}? 
    \item What if, in addition to observing only 0\% to 20\% of transitions and variable adoption of ORCID by gender, the transitions that were observed were heavily weighted toward STEM such that non-STEM usage rates were 0\% to 5\% while STEM usage rates were 15\% to 20\%? This scenario directly reflects the observations of Dasler and colleagues in their 2017 study of ORCID usage~\cite{dasler2017study}.
\end{enumerate}

Prior to providing more detail about the various methods for data censoring and biasing tested here, we introduce a simple method for creating synthetic data that is anchored in empirically observed values. Our ultimate goal in creating synthetic data is to generate data of the same form that we analyze in our regression analyses of field transitions (Main Model III)---that is, counts of women and men observed in transition from field $i$ to field $j$ ($W_{ij}$ and $M_{ij}$, respectively). To that end, let there be $30$ academic fields $i=1, 2, \dots, 30$, each with a field-specific covariate $x_i$ and a fraction of women $w_i$. Values of $w$ and $x$ are drawn at random. In particular, $x_i \sim \text{UNIFORM}[2,5]$ and $w_i \sim \text{UNIFORM}[0.15,0.85]$, IID. These ranges were meant to reflect ranges observed in empirical data. 

Our model proceeds by first stochastically choosing the total number of migrants from $i$ to $j$ of any gender, and then stochastically choosing whether each migrant is a man or a woman depending on a (possibly biased) function of parameters. First, let the total number of people moving from $i$ to $j$ be given by $N_{ij}$, an integer drawn from a geometric distribution with mean $\bar{N}$. Let each of these $N_{ij}$ people be a woman independently of all others with probability $p_{ij}$ and a man otherwise, according to the model 
\begin{equation}
    p_{ij} = \frac{1}{1+\exp[-\beta(x_j - x_i)-\log\tfrac{w_i}{1-w_i} - b_{ij}]}\ ,
\end{equation}
where $b_ij$ is a gender bias term indicating uneven sampling of women and men transitioning between field $i$ and field $j$. 

Note that when $\beta=0$ and $b_{ij}=0$, then $p_{ij} = w_i$. In other words, when there is no effect of the covariate $x$ and no gender bias in ORCID participation, the gender ratio among migrants from $i$ to $j$ is exactly the gender balance in the field of emigration $w_i$. Having computed values for $p_{ij}$, we then assign a simulated gender to each migrant, resulting in values for $W_{ij}$ and $M_{ij}$. 

We now show that, under increasingly extreme censoring and bias of the counts $M$ and $W$ (see Bias Scenarios 1 through 5 below), it is nevertheless possible to recover $\beta$, which measures the effect of the covariate $x$ on $M$ and $W$. In other words, we show that the model remains numerically consistent under known and speculated issues with ORCID data. 

Let {\bf Bias Scenario 1} be a simple test of consistency with copious and unbiased data. In plain language, Bias Scenario 1 is no bias at all, where we have ten times the amount of data under study. If the regression is unable to estimate the correct values in this scenario, the current study would be hopeless. 

We choose $\bar{N}$ so that there are, in expectation, $787,980$ observed transitions---a number that is ten times the amount of data available to us in our study. Then, for various true values of $\beta$, we ask whether we are indeed able to accurately estimate $\beta$ using the regression analysis described in the main text. Bias Scenario 1 is meant to simulate scenarios in which there is far more participation in ORCID across all fields. 

Let {\bf Bias Scenario 2} be identical to Bias Scenario 1, except that we now multiply the previous $\bar{N}$ by $0.1$ so that there are, in expectation, $78,798$ transitions observed. In plain language, Bias Scenario 2 imagines that ORCID users are a uniform 10\% sample of the much larger dataset used in Bias Scenario 1. The size of the dataset in Bias Scenario 2 matches the observed ORCID data. Again, for true values of $\beta$, we ask whether we are able to accurately estimate $\beta$ using the regression analysis described in the main text, but this time, using 10\% of the data. 

Let {\bf Bias Scenario 3} be similar to Bias Scenario 2, in that we target $78,798$ synthetically generated transitions, but instead of multiplying the original target counts $N_{ij}$ by $0.1$, we multiply them by a value chosen uniformly at random between $0$ and $0.2$, such that in expectation, again, one tenth of the transitions are removed, but this time they are removed heterogenously across fields. In plain language, Bias Scenario 3 imagines that ORCID users are a variable and noisy sample of the much larger dataset, where transitions between any two fields are observed at a rate chosen randomly between 0\% and 20\%.  

Let {\bf Bias Scenario 4} be identical to Bias Scenario 3 in terms of data censoring (with only $78,798$ observed transitions in expectation, and heterogeneous censoring of flows), but with a variably nonzero gender bias $b$. In particular, we draw $b_{ij}$ from a standard normal distribution $N(0,1)$ IID for each flow $i \to j$. For Bias Scenario 4, we independently repeat the process ten times.  

In plain language, Bias Scenario 4 imagines that ORCID users are a variable and noisy sample of the much larger dataset, where transitions are observed at a rate chosen randomly between 0\% and 20\%, just like Bias Scenario 3, but with an additional constraint. This scenario assumes that there is gender bias in reporting, such that one gender is more likely to have public-facing ORCID profiles than the other, with said biases drawn differently for each $i \to j$ flow.    

Finally, let {\bf Bias Scenario 5} be identical to Bias Scenario 4, but with a variable amount of censoring of the data to reflect higher ORCID users among those who are currently in, or have ever been in, STEM fields. Whereas in Bias Scenario 4 transitions were observed at a rate between 0\% and 20\% that was {\it independent} of the fields $i$ and $j$ involved in the $i \to j$ flow, here we let the rates of censoring depend on whether $i$ or $j$ is a STEM field. Specifically, if $i$ or $j$ is a STEM field, censoring rates were chosen uniformly between 15\% and 20\%, while if neither $i$ nor $j$ is a STEM field, censoring rates were chosen uniformly between 0\% and 5\%. 

\begin{figure*}
    \includegraphics[width=0.75\textwidth]{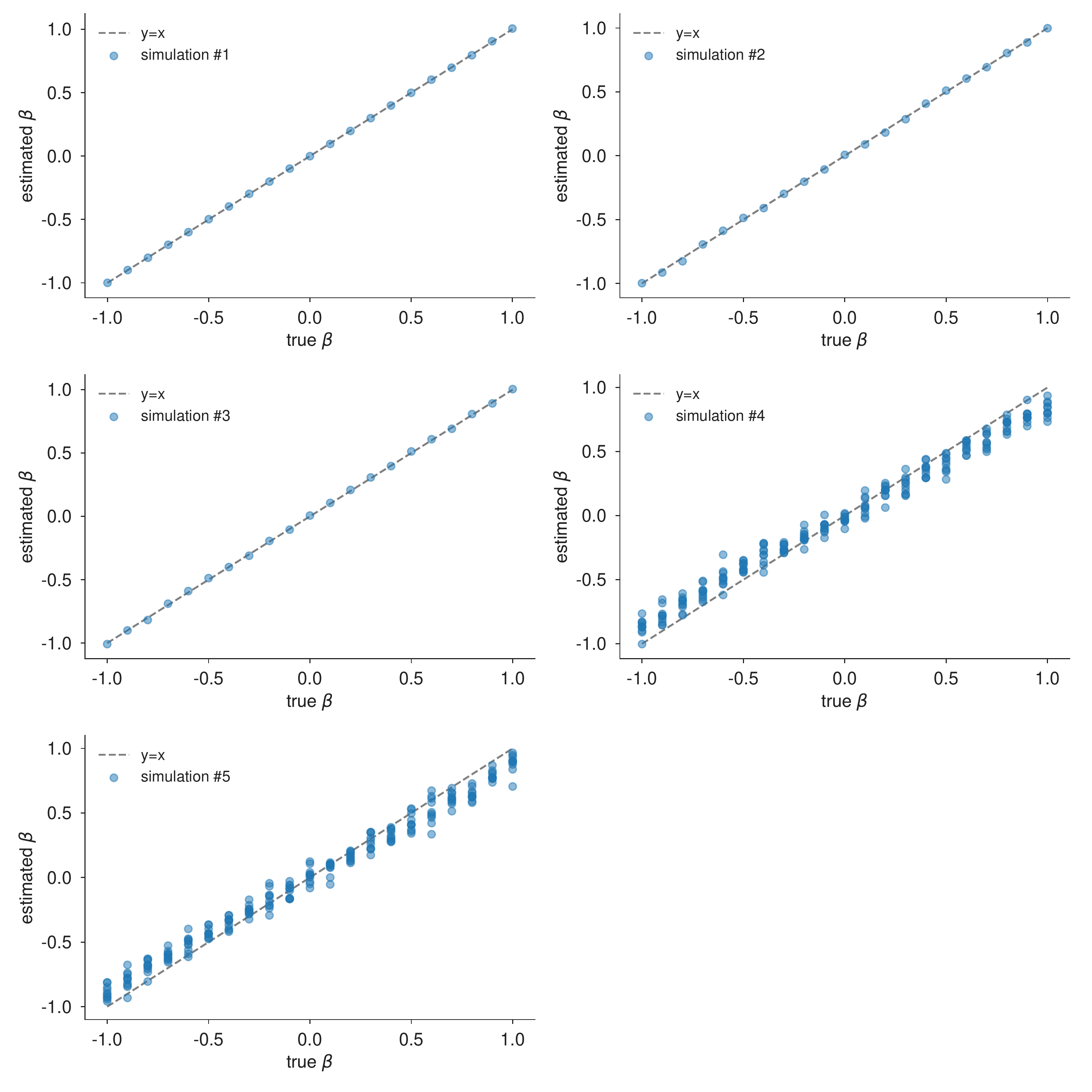}
    \captionsetup{justification=centering}
    \caption{Results of simulations for Bias Scenarios 1--5. In all cases, the estimated parameters are close to the true parameters.}
    \label{Fig:bias-scenario-simulations}
\end{figure*}

Across all five scenarios, and for a variety of choices of true $\beta$ and repeatedly redrawing $b_{ij}$ for ten technical replicates of Bias Scenarios 4 and 5, we find that the estimated $\beta$ values are a close match to the true $\beta$ values (see Figure~\ref{Fig:bias-scenario-simulations}). In other words, both heterogeneous censoring and variable gender and field sampling biases do not interfere with the regression's ability to accurately estimate effect sizes $\beta$. 

Intuitively, this robustness is due to the fact that the quantities being predicted in the regression---the relative probability that a migrant is a man or a woman---are stable to the censoring or subsampling of data and to overall levels of bias. Even if we only observe, say, one out of every 10 men and one out of every 20 women moving between fields, $\beta$ is nevertheless recoverable. 

Still, there remains the possibility that there are varieties of sampling bias that could affect our results, particularly when this bias is correlated with a covariate. For instance, if it is the case that making a transition between fields has a differential relationship with men's and women's choices to make a public-facing ORCID, which is further correlated with or magnified by a covariate like a field's brilliance orientation or selectivity, the regression models used in this study would not be able to identify and account for this form of bias. However, this bias scenario is unlikely. Under this scenario, for instance, it would have to be the case that more women than men who transition from physics (high brilliance orientation) to psychology (low brilliance orientation) {\it just so happen} to have a public-facing ORCID profile, while more men than women who transition from psychology to physics {\it just so happen} to have a public-facing ORCID profile. Although we cannot rule out this type of ``just so'' sampling bias, the probability of such systematic coincidences across all 870 possible $i \to j$ pairs of field transitions in this dataset is small.

\end{document}